\documentclass{aa}  

\usepackage{graphicx}
\usepackage{txfonts}
\usepackage[
breaklinks=true,
colorlinks=true,
linkcolor=black,
citecolor=blue,
urlcolor=blue
]{hyperref} %% to avoid \citeads line fills
\begin{document}

\title{Chemical abundances of stars with brown-dwarf companions}

%\subtitle{I. Overviewing the $\kappa$-mechanism}

\author{D. Mata S\'anchez\inst{\ref{inst1},\ref{inst2}} 
\and J. I. Gonz\'alez Hern\'andez\inst{\ref{inst1},\ref{inst2}} 
\and G. Israelian\inst{\ref{inst1},\ref{inst2}} 
\and N. C. Santos\inst{\ref{inst3},\ref{inst4}}
\and J. Sahlmann\inst{\ref{inst5},\ref{inst6}} 
\and S. Udry\inst{\ref{inst5}}
}

\institute{Instituto de Astrof\'isica de Canarias, 38200 La Laguna, Tenerife, 
Spain\label{inst1}\\
\email{dmata@iac.es,jonay@iac.es}
\and 
Departamento de Astrof\'isica, Universidad de La Laguna, 38206 La Laguna,
Tenerife, Spain\label{inst2} 
\and 
Centro de Astrof\'isica, Universidade do Porto, Rua das Estrelas, 4150-762
Porto, Portugal\label{inst3}
\and 
Departamento de F\'isica e Astronomia, Faculdade de C\^iencias, Universidade do
Porto, Portugal\label{inst4} 
\and 
Observatoire Astronomique de l'Universit\'e de Gen\`eve, 51 Ch. des Maillettes,
CH-1290 Sauverny, Versoix, Switzerland\label{inst5} 
\and
European Space Agency, European Space Astronomy Centre, P.O. Box 78,
Villanueva de la Ca\~nada, 28691 Madrid, Spain\label{inst6}
}

\date{Received March 13, 2014; accepted May XX, 2014}
 
\abstract
{It is well-known that stars with giant planets are on average more metal-rich
than stars without giant planets, whereas stars with detected low-mass planets do
not need to be metal-rich.}
{With the aim of studying the weak boundary that separates giant planets
and brown dwarfs (BDs) and their formation mechanism, we analyze the spectra of a sample 
of stars with already confirmed BD companions both by radial velocity and 
astrometry.}
{We employ standard and automatic tools to perform an EW-based analysis and
to derive chemical abundances from CORALIE spectra of stars with BD companions. 
}
{We compare these abundances with those of stars without detected planets and
with low-mass and giant-mass planets. 
We find that stars with BDs do not have 
metallicities and chemical abundances similar to those of 
giant-planet hosts
but they resemble the composition of stars with low-mass planets. The
distribution of mean abundances of $\alpha$-elements and iron peak elements of 
stars with BDs exhibit a peak at about solar abundance 
whereas for stars with low-mass and
high-mass planets the [X$_\alpha$/H] and [X$_{\rm Fe}$/H] peak 
abundances remain at $\sim -0.1$~dex and $\sim +0.15$~dex, 
respectively. 
We display these element abundances for stars with low-mass and 
high-mass planets, and BDs versus the minimum mass, $m_C \sin i$, 
of the most-massive substellar companion in each system, and we 
find a maximum in $\alpha$-element as well as Fe-peak abundances at 
$m_C \sin i \sim 1.35\pm 0.20$ jupiter masses.}
{We discuss the implication of these results in the context of the  
formation scenario of BDs in comparison with that of giant planets.}


\maketitle
%
%________________________________________________________________

\section{Introduction}

The most extended convention place the mass range of brown dwarfs
(BDs) at $ 13 - 80 \, M_{J}$ (being $M_J$ the mass of Jupiter), having 
enough mass to burn deuterium but not for hydrogen fusion~\citep{bur97},
i.e. in between the heaviest giant planets and 
the lightest stars. 
BDs above $\sim 65 \, M_J$ are thought to fuse lithium and
therefore the detection of the \ion{Li}{I}~$\lambda$~6708~{\AA} could be used 
to identify BDs, this is the so-called ``Lithium Test''~\citep{reb92}.

BDs were predicted by \citet{kum62} and \citet{hay63}, but they were not
empirically confirmed until 1995, when the first {\rm field} brown dwarf was 
detected \citep[\textbf{Teide 1},][]{reb95}. This occurs the same year as 
the discovery of the first extra-solar planet~\citep{may95}. 
The first BD companion to a M-dwarf
star was also discovered that year~\citep[\textbf{GJ 229B},][]{nak95}. 
During the following two decades high-precision radial velocity (RV) surveys 
have shown that close BDs around solar-type stars are 
rare~\citep[][and references therein]{gre06}. 
Thus, at orbital separations of less than 10 AU, the frequency
BD companions remains below 1~\%~\citep{mar00}, whereas it is $\sim 7$~\% for 
giant planets~\citep{udr07,may11} and $\sim 13$~\% for stellar 
binaries~\citep{hal03}. 
The so-called "Brown dwarf desert"
may be interpreted as the gap between the largest-mass objects that 
can be formed in protoplanetary discs, and the smallest-mass clumps that 
can collapse and/or fragment in the vicinity of a
protostar~\citep{mag14}. 
The mass function, $dN/dm_C \propto m_C^\alpha$, of close planetary and stellar 
companions drops away ($\alpha \sim -1$) towards the BD mass 
range~\citep{gre06}.
On the other hand, the mass function of isolated substellar objects is
roughly flat or even with linear increase ($\alpha\sim 0$) down to 
$\sim 20$~$M_J$~\citep{cha02,kir12}. This may point to a different formation 
scenario for close BD companions and BDs in the field and clusters. 

\citet{sah11} presented the discovery of nine BD companions from a sample
of 33 solar-type stars that exhibit RV variations caused by a companion 
in the mass range $m_C \sin i \sim 13-80$~$M_J$. They used Hipparcos 
astrometric data~\citep{per97} to confidently discard some of the BD 
candidates. Including literature data, these authors quoted 23 
remaining potential BD candidates. 
From CORALIE planet-search sample, they obtain an upper limit of 0.6\% 
for the frequency of BD companions around Sun-like stars.  
Recently, \citet{mag14} have collected all the BD candidates 
available in the literature including those in \citet{sah11}, 
some from the SDSS-III MARVELS survey~\citep{gej08} and some 
other RV surveys~\citep[e.g.][]{mar00}.

The metallicity of stars with BD companions have been briefly discussed in
\citep{sah11}. They note that the sample is still too small to claim any 
possible metallicity distribution of stars hosting BDs. 
\citet{mag14} extended the sample to roughly 65 stars with BD 
candidates, 
including dwarfs and giants, and stated that the mean metallicity of 
their sample is $\langle$[Fe/H]$\rangle$~$=-0.04$ ($\sigma=0.28$), i.e.
remarkably lower than that of stars with giant planets
($\langle$[Fe/H]$\rangle$~$=+0.08$, \citealp{sou08,sou11}). 
On the other hand,
stars with only detected ``small'' planets (hereafter ``small'' 
planet refers to a low-mass planet, including Super-Earths and
Neptune-like planets, with $m_C \sin i < 30 M_\oplus$,
whereas ``giant'' planet refers to high-mass planets, including
Saturn-like and Jupiter-like planets, with 
$30 M_\oplus < m_C \sin i < 13 M_J$, see Section~\ref{sec3}) 
do not seem to require high metal content to form planets within planetary 
discs~\citep{sou08,sou11,adi12b}. 
\citet{sou11} study a sample of 107 stars with planets (97 giant and 
10 small planets) 
and found an average metallicity of stars with small planets at about 
$\langle$[Fe/H]$\rangle$~$=-0.11$, very similar to that of stars without 
detected planets~\citep{sou08}. 

Currently, there are two well-established 
theories for giant planet formation: core-accretion scenario~\citet{pol96} 
and disc gravitational instability~\citep{bos97}. 
The core-accretion model is more sensitive to the 
fraction of solids in a disc than is the disc-instability model. 
The formation of BDs has been also extensively studied. Two main 
mechanism have been proposed: molecular cloud 
fragmentation~\citep{pad04}, and disc fragmentation~\citep{sta09}. 
The latter mechanism, which requires a small fraction of Sun-like stars 
should host a massive extended disc, is able to explain most 
of the known BDs which may either remain bound to the primary star, 
or be ejected into the field~\citep{sta09}.  

In this paper, we present a uniform spectroscopic analysis for a sample of
stars with BD companions from \citet{sah11} and we compare the results with 
those of a sample of stars with known giant and small planets from previous 
works~\citep{adi12b}. The aim of this work is to provide some
information that could be useful to distinguish among the different and 
possible formation mechanisms of BD companions. 

%__________________________________________________________________

\begin{table*}

\caption[]{Stellar parameters of the CORALIE sample}
\label{tpar}
\centering
\begin{tabular}{lcccccc}
\noalign{\smallskip}
\noalign{\smallskip}
\noalign{\smallskip}
\hline\hline
\noalign{\smallskip}
Star & $ T_{\rm eff}$ & $ \log g$   & $ \xi_t$ & $\rm [Fe/H]$ & $ M_2 \sin i$	    & References \\ 
    & $\rm [K]$ & $\rm [dex]$ & $\rm [cm/s]$ & $\rm [dex]$ & $ [M_J]$
\\
\hline
\noalign{\smallskip}
HD4747       & $5316\pm 50$ & $4.48\pm 0.10$ & $0.79\pm 0.10$ & $-0.21\pm 0.05$ & $46.1$ & 2\\ 
HD52756      & $5216\pm 65$ & $4.47\pm 0.11$ & $1.11\pm 0.13$ & $0.13\pm 0.04$ & 59.3  & 1\\   
HD74014      & $5662\pm 55$ & $4.39\pm 0.08$ & $1.10\pm 0.07$ & $0.26\pm 0.04$ & 49.0 & 1\\ 
HD89707      & $6047\pm 42$ & $4.52\pm 0.05$ & $0.99\pm 0.06$ & $-0.33\pm 0.03$ & 53.6  & 1\\     
HD167665     & $6224\pm 39$ & $4.44\pm 0.04$ & $1.18\pm 0.05$ & $-0.05\pm 0.03$ & 50.6    & 1\\  
HD189310     & $5188\pm 50$ & $4.49\pm 0.09$ & $0.94\pm 0.10$ & $-0.01 \pm 0.03$ & 25.6 & 1 \\ 
HD211847     & $5715\pm 24$ & $4.49\pm 0.05$ & $1.05\pm 0.03$ & $-0.08\pm 0.02$ & 19.2 & 1  \\ 
\noalign{\smallskip}
\hline
\noalign{\smallskip}
HD3277$^a$   & $5539\pm 49$ & $4.36\pm 0.06$ & $0.91\pm 0.07$ & $-0.06\pm 0.04$ & 64.7 & 1\\ 
HD17289$^a$  & $5924\pm 32$ & $4.37\pm 0.04$ & $1.15\pm 0.04$ & $-0.11\pm 0.03$ & 48.9 & 1\\ 
HD30501$^a$  & $5223\pm 27$ & $4.56\pm 0.08$ & $1.18\pm 0.04$ & $-0.06\pm 0.02$ & 62.3 & 1\\ 
HD43848$^a$  & $5334\pm 92$ & $4.56\pm 0.15$ & $1.35\pm 0.17$ & $0.22\pm 0.06$ & 24.5 & 1\\ 
HD53680$^a$  & $5167\pm 94$ & $5.37^b\pm 0.29$ & $2.08\pm 0.31$ & $-0.29\pm 0.04$ & 54.7  & 1\\  
HD154697$^a$ & $5648\pm 45$ & $4.42\pm 0.05$ & $1.04\pm 0.06$ & $0.13\pm 0.04$ & 71.1 & 1\\ 
HD164427A$^a$& $6003\pm 27$ & $4.35\pm 0.03$ & $1.19\pm 0.03$ & $0.19\pm 0.02$ & 48.0   & 1\\     
HIP103019$^a$& $4913\pm 115$& $4.45\pm 0.28$ & $0.54^c\pm 0.10$ & $-0.30\pm 0.06$ & 52.5 & 1\\ 
\noalign{\smallskip}
\hline
\noalign{\smallskip}
HD74842$^d$  & $5517\pm 38$ & $4.50\pm 0.06$ & $1.01\pm 0.06$ & $-0.08\pm 0.03$ & -- & 3 \\ 
HD94340$^d$  & $5902\pm 26$ & $4.19\pm 0.03$ & $1.30\pm 0.03$ & $0.11\pm 0.02$ & -- & 3 \\ 
HD112863$^d$ & $5342\pm 36$ & $4.57\pm 0.07$ & $1.08\pm 0.07$ & $-0.11\pm 0.03$ & -- & 3\\ 
HD206505$^d$ & $5392\pm 44$ & $4.46\pm 0.07$ & $1.02\pm 0.07$ & $0.11\pm 0.03$ & -- & 3\\ 
\noalign{\smallskip}
\hline   
\noalign{\smallskip}
\noalign{\smallskip}
\end{tabular}
\tablebib{(1) \citet{sah11}; (2) \citet{san05}; (3) This work.}
\tablefoot{\tablefoottext{a}{These eight stars have companion minimum masses, 
$m_C \sin i$, in the BD range determined from spectroscopic RV 
measurements, but are discarded in \citet{sah11}, from their 
Hipparcos astrometry.}\\
\tablefoottext{b}{The surface gravity of the star HD~53680 is
unusually for its  derived effective temperature. A significantly 
lower $T_{\rm eff}$ value (probably $ < 4500$~K) is expected from its 
weak and narrow $H\alpha$ profile (see Fig.~\ref{fspec} and 
Section~\ref{secabu}).}\\
\tablefoottext{c}{The microturbulence of HIP~103019 was calculated 
following the expression presented in \citet{adi12c}.}\\
\tablefoottext{d}{These four stars, as a comparison sample, are also 
from the CORALIE sample but they do not have detected BD companions}
}
\end{table*}

\section{Observations\label{secobs}}

We analyse data for two different samples obtained with two different 
telescopes and instruments: stars with BDs with spectroscopic data 
at resolving power $R\sim 50,0000$ taken at the 1.2m-Euler Swiss Telescope 
equipped with the CORALIE spectrograph~\citep{udr00a} and stars with 
planetary companions observed with the HARPS 
spectrograph~\citep{may03} with $R\sim115,000$ installed at the 
3.6m-ESO telescope, both of them at 
La Silla Observatory (ESO) in Chile. 

The individual spectral of each star were reduced in a standard manner, and
latter normalized within the package IRAF\footnote{IRAF is distributed by
National Optical Astronomy Observatories, operated by the Association of
Universities for Research in Astronomy, Inc., under contract with the National
Science Foundation.}, using low-order polynomial fits to the observed 
scontinuum.

%__________________________________________________________________

\begin{figure}
\resizebox{\hsize}{!}{\includegraphics[angle=90]{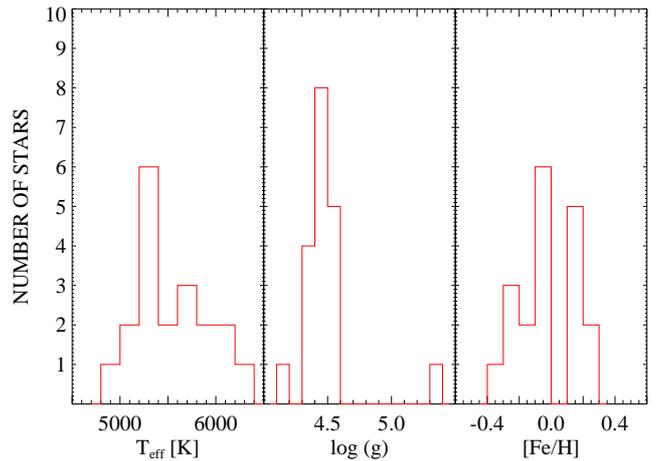}}
\caption{Histograms of the stellar parameters $T_{\rm eff}$, $\log{g}$ 
and $\rm [Fe/H]$ of our CORALIE sample.}
\label{fma1}
\end{figure}  

\section{Sample description and stellar parameters\label{sec3}}

\subsection{Stars with BD-companion candidates}

Our stellar sample has been extracted mostly from F-, G- and K-type 
main-sequence stars of the CORALIE RV survey~\citep{udr00a}.
This sample consists of 15 stars with BD companion candidates 
reported in~\citet{sah11},  
for which the minimum mass, $m_C \sin i$, of most massive companion is 
in the brown-dwarf mass range ($13 - 80\,  M_J$). 
One of these 15 stars, HIP~103019, has been extrated from the HARPS RV 
survey~\citep{may03}.
In Table~\ref{tpar} we provide the minimum mass of these 15 BD 
candidates.
\citet{sah11} were also able to derive the orbital inclination, $i$, 
by using astrometric measurements from Hipparcos~\citep{per97,van07}. 
This allowed them to confidently exclude as BD candidates eight stars 
from the initial sample of 15 stars because the current mass 
determinations, $m_C$, place them in M-dwarf stellar regime.
Stellar parameters of the sample of 14 stars were 
collected from \citet{sah11} and one star from \citet{san05}. 
Four additional stars from the CORALIE sample without detected BD 
companions were analyzed as a comparison/control sample 
(see Table~\ref{tpar}). 
In Fig.\ref{fma1} we show the histograms of $ T_{\rm eff}$, $\log g$, and 
[Fe/H] for our stellar sample.
We only display those stars with available $m_C \sin i$ values in 
Table~\ref{tpar}. 

\begin{figure}
\resizebox{\hsize}{!}{\includegraphics{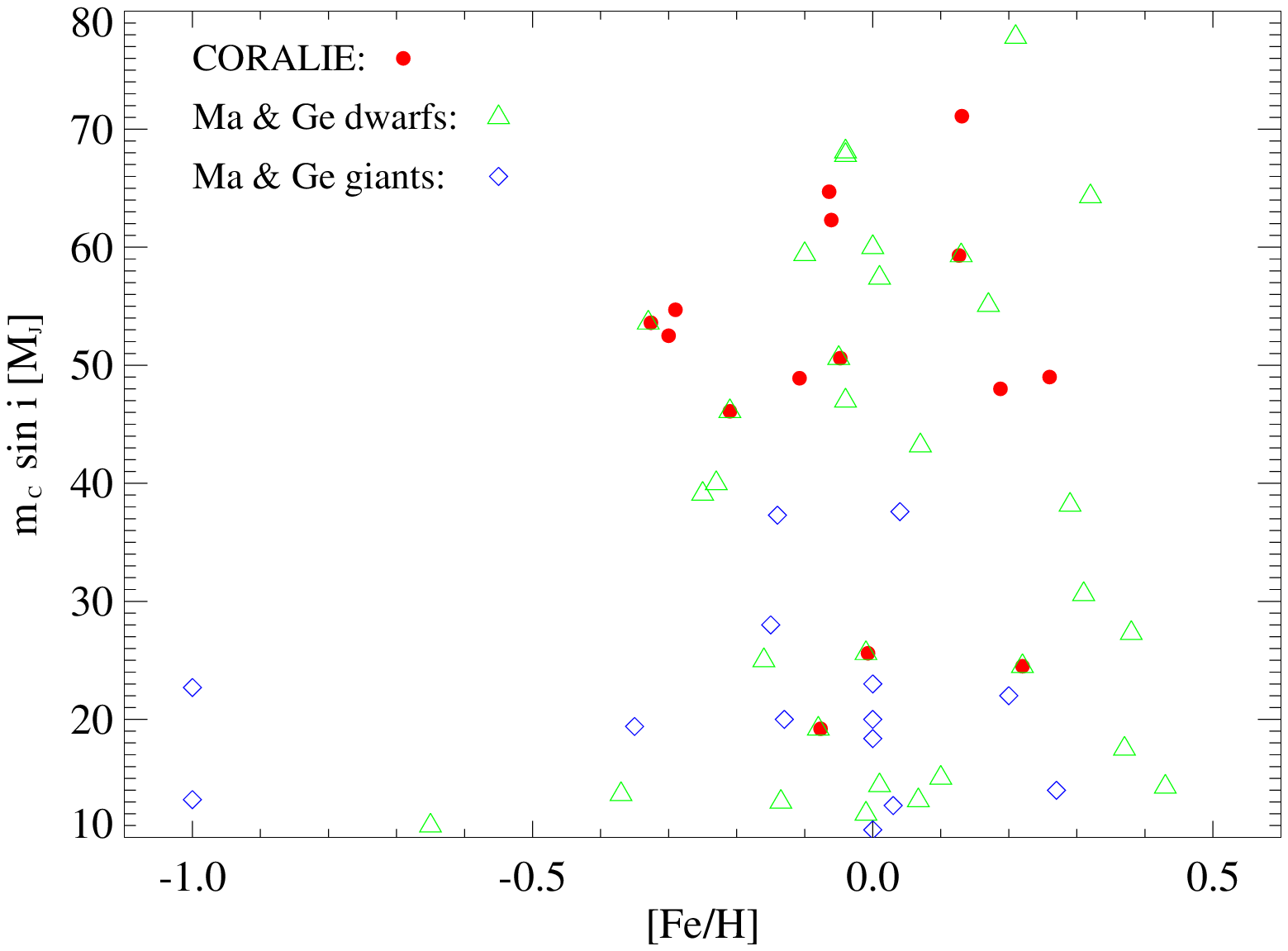}}
\resizebox{\hsize}{!}{\includegraphics{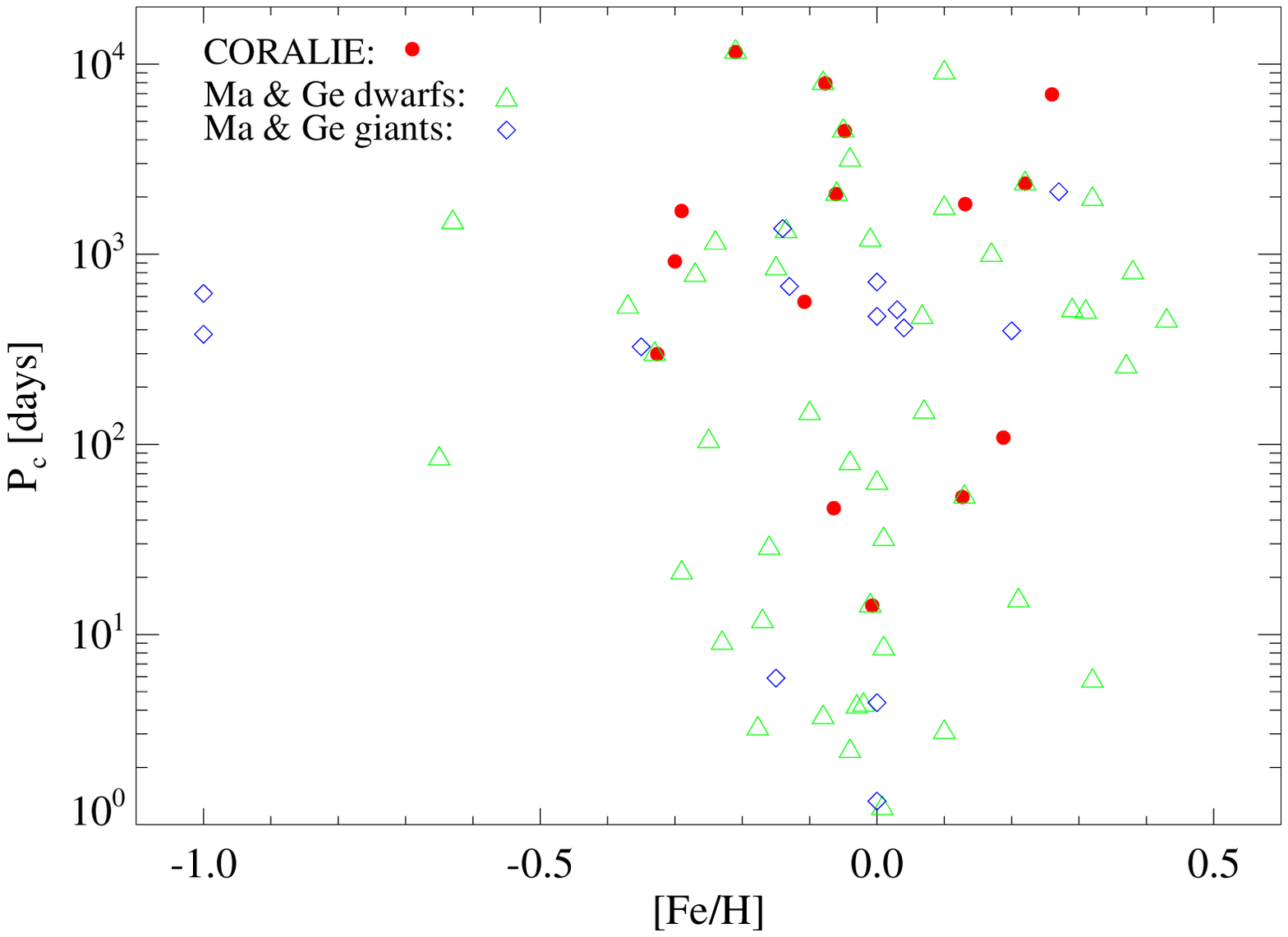}}
\caption{Minimum mass of the most-massive companion, $m_C \sin i$, and 
orbital period, $P_C$, versus the metallicity of stars with BDs. 
The CORALIE sample is depicted as filled symbols, and the sample 
in \citet{mag14} is displayed as empty symbols, separated among giant
and dwarf stars.}
\label{fma2}
\end{figure}

We note that \citet{mag14} also collected a sample 65 stars with BD
candidates: 43 stars, 27 dwarfs and 15 giants, with $m_C \sin i$ values 
in the range 13-90~$M_J$. 
Two stars, included in~\citet{mag14} as BD candidates, HD~30501 and 
HD~43848, were discarded by \citet{sah11}, probably because they have 
$m_C$ values above but close to the 80~$M_J$ boundary.
Fig.\ref{fma2} depicts the minimum mass of the BD companion $m_c \sin i$ 
and orbital period against metallicity [Fe/H], including stars of our 
sample as well as those stars in the sample of \citet{mag14} for comparison.
We separate giant ($ \log g < 4$) and dwarfs ($ \log g > 4$) stars. 
The stars with BD candidates spread over a wide range of orbital
periods, a minimum masses of the BDs, and stellar metallicities.

\subsection{Stars with planetary-mass companions}

The HARPS sub-sample \citep[HARPS-1 sample in][]{adi12b} used in this work 
contains 451 stars~\citep{sou08,nev09}, both with and without 
planetary companions. 
We collect the minimum mass of the most-massive planet in each 
planetary system from the encyclopaedia of extra-solar 
planets~\footnote{\url{http://exoplanet.eu}}. 
The planetary-mass sample is separated in two groups: 
(i) {\it small planets} (SP; super-Earth like and Neptune like planets) with 
masses of $m_C \sin i < \sim 0.094$~$M_J$ ($\sim 30 M_\oplus$), and 
(ii) {\it giant planets} (GP; Saturn like and Jupiter like planets) with 
masses in the range $0.094 < m_C \sin i \;[M_J] < 13$.
Two of the stars with giant planets, HD~162020 and HD~202206, within the HARPS 
sample have companion masses above 13~$M_J$ and will be considered as BDs 
hereafter. 

Therefore, our final sample of confirmed BDs contains 9 dwarf stars, with 
companions in the mass range $m_C \sin i \sim 13-80$~$M_J$.
In the following, we may refer to ``BD-host stars'' to stars with 
confirmed BDs, i.e. with $m_C \sim 13-80$~$M_J$, and ``stars with discarded 
BDs'' to those with $m_C \sin i \sim 13-80$~$M_J$ but $m_C > 80$~$M_J$, 
according to \citet{sah11}.
The sample of planet-host stars contains 25 stars
with small planets and 78 stars with giant planets. In Table~\ref{tplm} 
we provide the minimum mass of the most-massive planet in each planetary 
system of the stars in the HARPS sample. 

%\newpage

%__________________________________________________________________

\section{Automatic codes for EW measurements: ARES versus TAME.\label{secew}}

\begin{figure*}
\resizebox{\hsize}{!}{\includegraphics[angle=90]{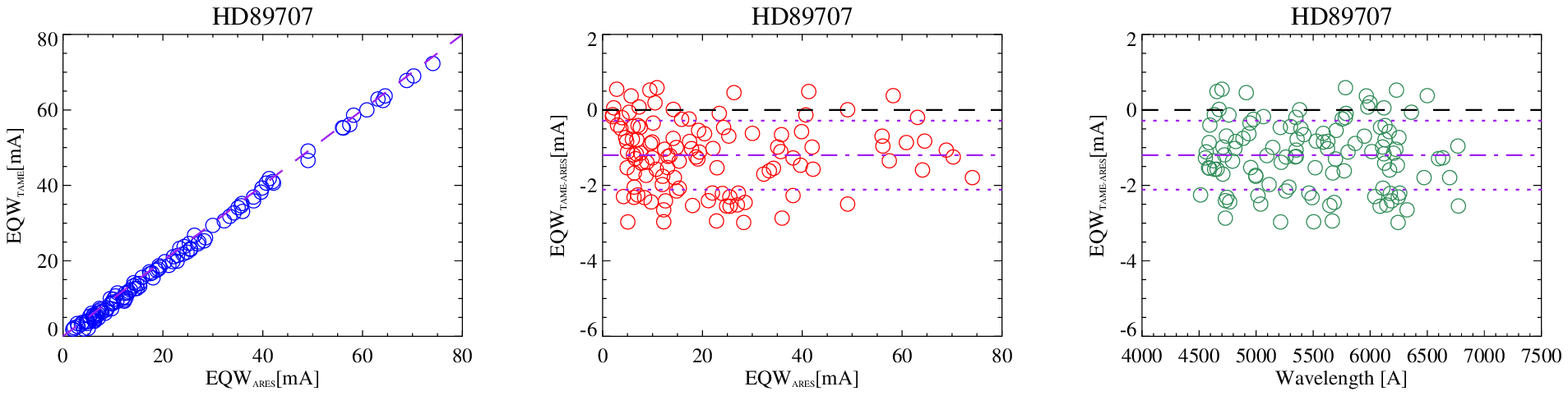}}
\resizebox{\hsize}{!}{\includegraphics[angle=90]{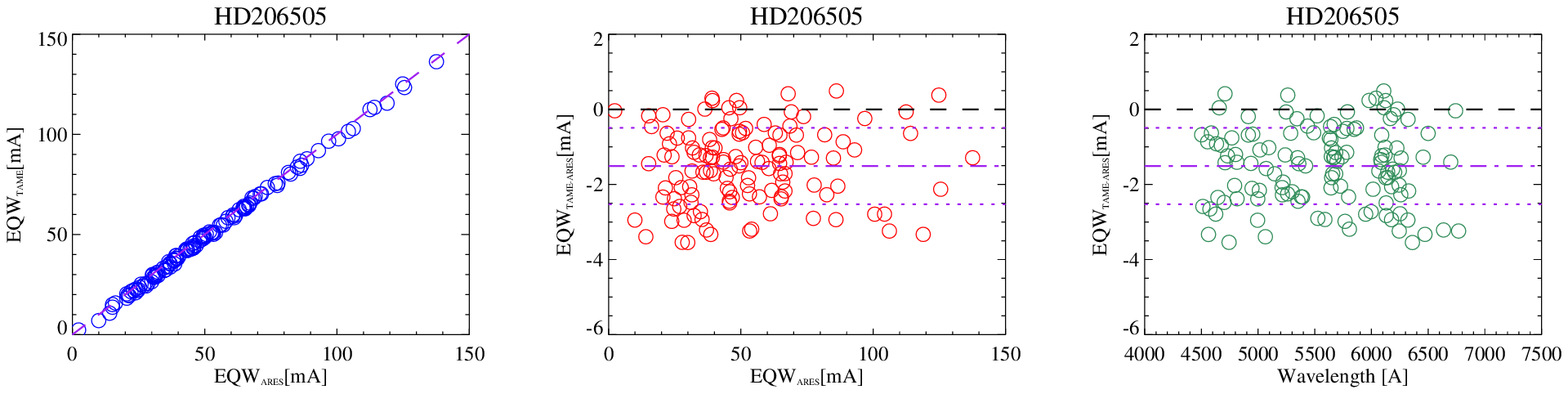}}
\caption{ARES versus TAME graphics for the stars HD~89707 
($\mathrm{S/N}\sim110$) and HD~206505($\mathrm{S/N}\sim70$).
{\it Left panels}: EW measured with TAME versus ARES. 
The 1:1 correspondence is shown as a dashed line.  
EW differences, $\rm EW_{TAME}-EW_{ARES}$, versus $\rm EW_{ARES}$ 
({\it middle panels}) and spectral line wavelength ({\it right panels}). 
Dash-dotted lines define the mean value of the data points, and dotted lines 
define the mean plus the standard deviation.
}
\label{few}
\end{figure*} 

\begin{figure}
\resizebox{\hsize}{!}{\includegraphics{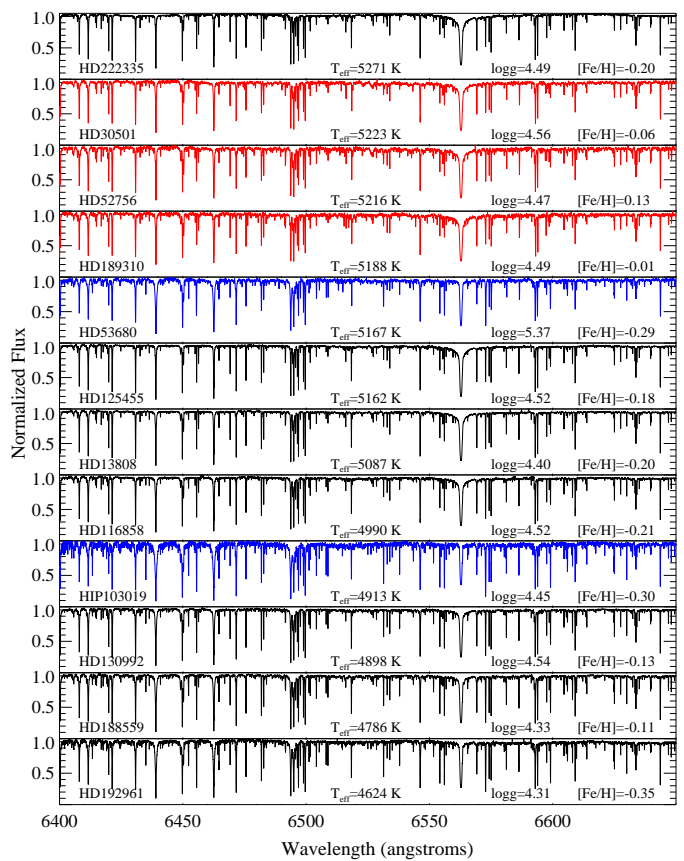}}
\caption{High-resolution normalized spectra of late G-, K-type stars: 
CORALIE spectra of coolest stars in the sample (black), HARPS spectra 
(red) of cool stars from the HARPS database~\citep[e.g.][]{sou08} and 
the CORALIE spectrum (blue) of HD~53680 and the HARPS spectrum (blue) 
of HIP~103019. 
The spectra are depicted following a sequence with decreasing 
$T_{\rm eff}$ from top to bottom.}
\label{fspec}
\end{figure}

We measure the equivalent widths (EWs) of spectral lines using the 
linelists in \citet{sou08,nev09,adi12b} using automatic tools.
We explore two different automatic codes for EWs spectra analysis: the 
automatic C++ based code ARES\footnote{The ARES code can be downloaded at:\\
\url{http://www.astro.up.pt/}} \citep{sou07} and a new IDL based code named
TAME\footnote{The TAME code can be downloaded at:\\ 
\url{http://astro.snu.ac.kr/~wskang/tame/}} \citep{kan12}.
In order to compare these two automatic codes, we measure the EWs of CORALIE 
sample with the same input parameters to these two codes. 

In Fig.~\ref{few}, we compare the EWs measured using TAME, $EW_{\rm TAME}$, 
against those estimated using ARES, $EW_{\rm ARES}$.
The mean value of the EW differences $EW_{\rm TAME}-EW_{\rm ARES}$ is found 
at $\sim -1.2\, m\AA$ and $\sim -1.5\, m\AA$ for the stars HD~89707 star 
($\mathrm{S/N}\sim110$) and HD~206505($\mathrm{S/N}\sim70$). 
The TAME code is very slightly underestimating the EW compared to ARES 
measurements, and the scatter of these comparisons is lower than 
$\sim 1.5 \, \rm m\AA$. 
These EW differences do not exhibit any remarkable dependence on wavelength. 
We also tested whether the signal-to-noise ratio (S/N) from our stellar spectra 
is the source of this observed tendency and no trend has been found. 
The mean value of the EW differences is fluctuating in the 
$-2\, \rm m\AA$ to $-1\, \rm m\AA$ range. 
The standard deviation of the EW differences improves slightly as the 
S/N increases, but it oscillates between $0.5\,\rm m\AA$ to 
$1.5\,\rm m\AA$. 

This analysis lead us to conclude that both programs show a good agreement 
and their differences are not significant and do not have any relevant impact 
on the chemical abundance analysis, within the typical error bars of the
EW-based chemical abundance analysis.

\begin{table}
\caption{\ion{Fe}{i} and \ion{Fe}{ii} abundances and standard deviations}
\label{tafe}
\centering
\begin{tabular}{lrr}
\noalign{\smallskip}
\noalign{\smallskip}
\noalign{\smallskip}
\hline
\hline
\noalign{\smallskip}
Star & $\rm[Fe I/H]$ & $\rm[Fe II/H]$ \\ 
\\
\hline
\noalign{\smallskip}
    HD4747         & $-0.28  \pm 0.06  $ &    $-0.28  \pm 0.10   $ \\
   HD52756         & $ 0.09  \pm 0.10  $ &    $ 0.03  \pm 0.15   $ \\
   HD74014  	   & $ 0.23  \pm 0.07  $ &    $ 0.16  \pm 0.11   $ \\
   HD89707  	   & $-0.35  \pm 0.09  $ &    $-0.40  \pm 0.12   $ \\
  HD167665  	   & $-0.11  \pm 0.10  $ &    $-0.09  \pm 0.10   $ \\
  HD189310  	   & $-0.03  \pm 0.10  $ &    $-0.10  \pm 0.22   $ \\
  HD211847  	   & $-0.10  \pm 0.06  $ &    $-0.11  \pm 0.09   $ \\
\noalign{\smallskip}
\hline
\noalign{\smallskip}
    HD3277         & $-0.10  \pm 0.05  $ &    $-0.12  \pm 0.08   $ \\
   HD17289  	   & $-0.13  \pm 0.08  $ &    $-0.09  \pm 0.13   $ \\
   HD30501         & $-0.09  \pm 0.10  $ &    $-0.14  \pm 0.18   $ \\
   HD43848         & $ 0.18  \pm 0.13  $ &    $ 0.14  \pm 0.25   $ \\
  HD53680$^\star$  & $-0.37  \pm 0.24  $ &    $-0.61  \pm 0.41   $ \\
  HD154697  	   & $ 0.10  \pm 0.06  $ &    $ 0.06  \pm 0.10   $ \\
 HD164427A  	   & $ 0.15  \pm 0.06  $ &    $ 0.14  \pm 0.09   $ \\
HIP103019$^\star$  & $-0.34  \pm 0.29  $ &    $-0.68  \pm 0.45   $ \\
\noalign{\smallskip}
\hline
\noalign{\smallskip}
   HD74842  	   & $-0.12  \pm 0.09  $ &    $-0.10  \pm 0.11   $ \\
   HD94340  	   & $ 0.10  \pm 0.06  $ &    $ 0.08  \pm 0.11   $ \\
  HD112863  	   & $-0.12  \pm 0.09  $ &    $-0.14  \pm 0.10   $ \\
  HD206505  	   & $ 0.10  \pm 0.09  $ &    $ 0.08  \pm 0.12   $ \\
\noalign{\smallskip}
\hline   
\noalign{\smallskip}
\noalign{\smallskip}
\end{tabular}
\tablefoot{\tablefoottext{$\star$}{These stars were discarded for the 
abundance analysis due to the large scatter on the $\ion{Fe}{i}$ and 
$\ion{Fe}{ii}$ abundances which may be related to the fact that these 
stars are probably not well classified and may have in fact 
lower $T_{\rm eff}$ (see Section~\ref{secabu})}
} 
\end{table}

%__________________________________________________________________

\section{Chemical abundances\label{secabu}}

We compute the EWs using ARES for consistency with chemical abundance 
analysis in \citet{adi12b}. We also use version 2010 of the 
MOOG~\footnote{The MOOG code can be 
downloaded at: \url{http://www.as.utexas.edu/~chris/moog.html}} 
code~\citep{sne73} together with Kurucz ATLAS9 stellar model 
atmospheres~\citep{kur93} for chemical abundance determination. 

We first check the \ion{Fe}{i} and \ion{Fe}{ii} abundances, using the line 
list from \citet{sou08}. The high dispersion, $\sigma$, of the Fe 
abundances (see Table~\ref{tafe}) of the stars HD~53680 and 
HIP~103019 suggests that these stars may have lower effective 
temperatures than those given Table~\ref{tpar}. 
In Fig.~\ref{fspec} we depict the normalized spectra of the coolest
stars in the sample together with some spectra of late-G, K-type dwarfs from
the HARPS database~\citep[e.g.][]{sou08}. 
The $H\alpha$ profiles of these
two stars do not follow the sequence of temperatures but they 
appear to be the coolest objects in Fig.~\ref{fspec}. 
In addition, these stars were discarded as candidates of BDs 
by \citet{sah11}. We note the unusually high surface gravity
estimated for the star HD~53680 which is likely spurious, consistent
with the large scatter in the Fe abundances and the difference 
between \ion{Fe}{i} and \ion{Fe}{ii} abundances. This star is
catalogued in the SIMBAD database as a K6V star in a visual binary
system. The narrow $H\alpha$ line and the large dispersion in 
\ion{Fe}{i} and \ion{Fe}{ii} abundances may indicate an even later 
spectral type.
For these reasons, they may deserve further analysis and 
from this point on, we will not consider them.

We use the linelist in \citet{nev09} on 17 stars of CORALIE remaining sample
(seven stars with confirmed BDs, six with discarded BD candidates, 
and four without detected BD companions)
to derive element abundances of \ion{Na}{i}, \ion{Mg}{i}, \ion{Al}{i},
\ion{Si}{i}, \ion{Ca}{i}, \ion{Sc}{i}, \ion{Sc}{ii}, \ion{Ti}{i},\ion{ Ti}{ii}, 
\ion{V}{i}, \ion{Cr}{i}, \ion{Cr}{ii}, \ion{Mn}{i}, \ion{Co}{i} and \ion{Ni}{i}.
Some lines of the original list are not considered: lines non-detected by 
ARES, lines discarded in the \citet{adi12b}, and lines whose abundances 
were out of the $3\sigma$ range. 
The abundance results are shown in Tables~\ref{tabu1},~\ref{tabu2} 
and~\ref{tabu3}. 
The chemical abundances of the HARPS sample were obtained from 
\citet{adi12b} where they use exactly the same tools and model atmospheres. 

%__________________________________________________________________

\section{Discussion\label{secdisc}}

\citet{gon97, gon00} already noticed that giant planet hosts tend to be
more metal-rich than stars without detected planets. 
\citet{san01} provided supporting evidences of a metal-rich origin of
giant-planet host stars and following studies confirmed this result 
(e.g.~\citealt{san04,san05,val05}). 
Recent studies show that Neptune and super-Earth class planet hosts have a 
different metallicity distribution, more similar to stars without planets 
(e.g. \citealp{udr06}; \citealp{sou08,sou11}; \citealp{ghe10}; 
\citealp{may11}; \citealp{buc12}). 

In this section, we inspect the abundance ratios of different elements, 
$\rm [X/Fe]$, as a function of the metallicity for our BD-companion stellar 
sample, as well as comparing them with the planetary-companion stars 
analized by \citet{adi12b}. 
We also study the distributions of different element 
abundances in different samples and we compare them with other stars with
and without planets as a function of the minimum mass of the most-massive 
companion of each host star, $m_C \sin{i}$. 

\subsection{Galactic abundance trends}

The abundances of the refractory elements in CORALIE sub-sample
exhibit the similar behaviour as other stars with and without planets 
analysed in previous works~\citep{nev09,adi12b}  
(see Fig.~\ref{ftrend}). 

Stars with BDs follow the galactic abundance trend except for some particular 
elements (Co, Si, Sc) where the abundances are slightly lower than expected. 
These exceptions may be due to the small number of lines to achieve a
reliable mean abundance in certain elements (e.g. ScI for HD~167665). 
Stars with BD companions appear to be located at an intermediate range of 
metallicity, between stars with and without planets. 

\begin{figure}
\resizebox{\hsize}{!}{\includegraphics{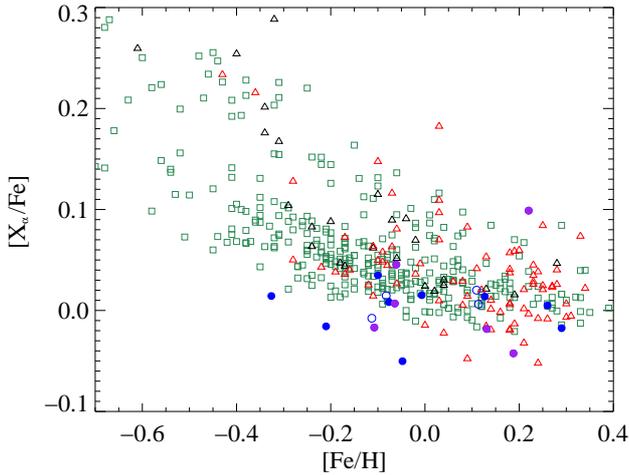}}
\caption{$\alpha$-element abundance ratios $[{\rm X_\alpha/Fe}]$ 
against [Fe/H] for the samples of stars without planets (green empty squares),
with small planets (black empty triangles), with giant 
planets (red empty triangles), BD-host stars (blue filled circles),
stars without known BD companions from the CORALIE sample (blue empty circles), 
and stars with discarded BD candidates (violet filled circles).}
\label{falpha}
\end{figure}

In Fig.~\ref{falpha} we display the mean abundance ratio of the 
$\alpha$-elements Mg, Si, Ca and Ti (with
[X$_\alpha$/H] computed as the sum of individual element abundances 
[X/H] divided by 4, and [X$_\alpha$/Fe]=[X$_\alpha$/H]-[Fe/H])
against the metallicity of stars with small and giant planets 
from~\citep{adi12a} together with stars with BDs.
The range of metallicities of stars with confirmed BDs seems 
to be narrower than that of the planet hosts although this may be 
not statistically significant due to the small number of stars in 
the BD sample.
\citet{adi12a,adi12c} remarked that stars with both small and giant 
planets at low metallicities [Fe/H]~$< -0.3$~dex, tend to be 
$\alpha$-enhanced and therefore to belong chemically to thick-disk 
population.
There is only one star with BD at these relatively low metallicities, at
[Fe/H]~$\sim -0.35$~dex, showing a relatively low [X$_\alpha$/Fe] ratio, and
therefore the question whether this behaviour still holds for BD hosts 
remains open.
The $\alpha$-element abundance ratios $[{\rm X_\alpha/Fe}]$ of BD hosts 
seem to be consistent with the Galactic trend at higher metallicities, 
although there are some stars at metallicities below solar with relatively 
low [X$_\alpha$/Fe] ratios, even below the trend described by the stars 
without detected planets of the HARPS sample. Stars with discarded BDs 
and without detected BD candidates also follow the general abundance 
trend.

\subsection{Element abundance distributions}

The histograms displayed in Fig.~\ref{fha} allow us to study the abundance
distribution [X/H] of our sample for each element. 
Stars without planets have a
maximum abundance at $ \sim -0.1$ for most speciess. 
Stars with giant planets exhibit a more metal-rich maximum at 
$\sim 0.2-0.3$~dex, while the stars with small planets (whose maximum 
is at $\sim -0.1$~dex) resemble to the ``single'' ones. 
This general behaviour already noticed in previous papers~\citep{nev09,adi12a},
is in a good agreement with the so-called metallicity effect, i.e. the 
strong correlation between stellar metallicity, and the likelihood of finding 
giant planets~\citep{san01}.

The abundance distribution of the sample of confirmed BD-host stars appears to
be located in between the stars with small planets and stars with giant 
planets.     
In fact, some element distributions (Na, Si, Mg, Mn) are more similar to those
of giant planets, whereas for others elements (Ti, Cr, Co, Ni) the behaviour 
is closer to the stars without planets (see Fig.~\ref{fha}). 

\begin{figure}
\resizebox{\hsize}{!}{\includegraphics{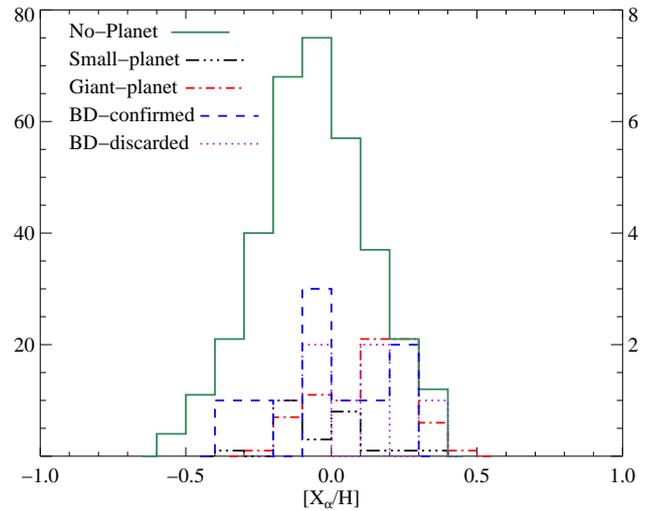}}
\resizebox{\hsize}{!}{\includegraphics{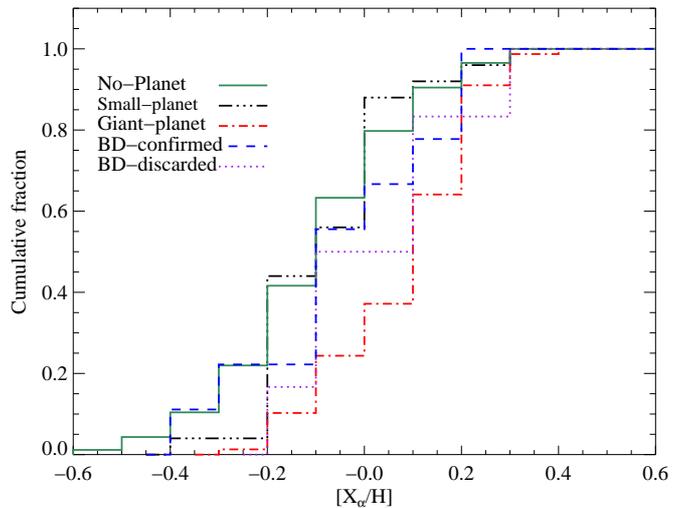}}
\caption{Histogram of $\alpha$-element abundances, $[{\rm X_\alpha/H}]$, 
(top panel) of the samples of stars without planets (green continuum 
line), stars with small planets (black dashed-double-dotted line), 
stars with giant planets (red dashed-dotted line), stars with 
confirmed BD companions (blue dashed line) and stars with discarded 
BD candidates, i.e. binaries with low-mass M-dwarf companions 
(violet dotted line). 
The left y-axis of the top panel is labeled with the number of stars 
with and without small and giant planets whereas the right y-axis 
shows the number of stars with BD-companion candidates.
The lower panel shows cumulative histogram.} 
\label{fhalpha}
\end{figure} 

\begin{figure}
\resizebox{\hsize}{!}{\includegraphics{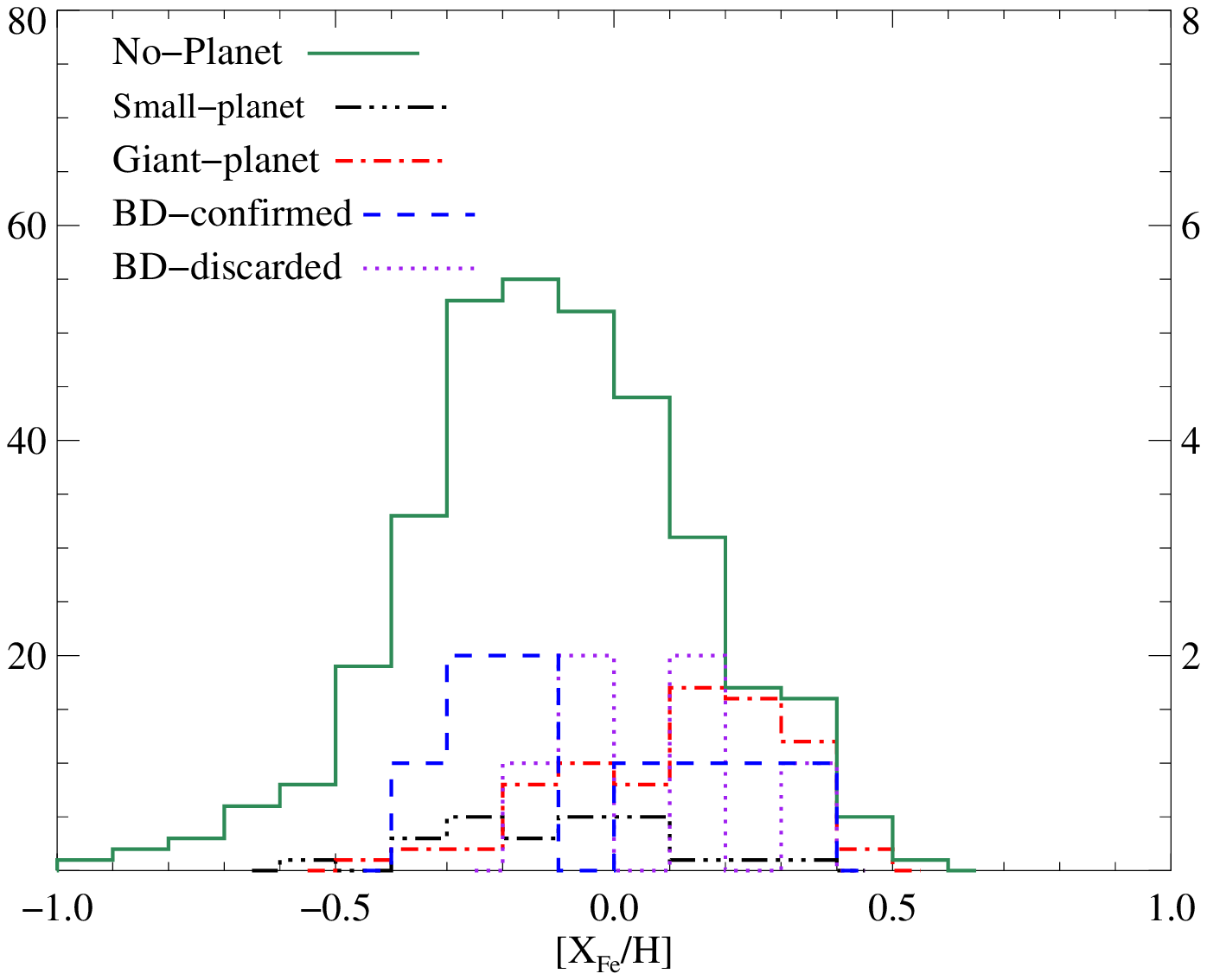}}
\resizebox{\hsize}{!}{\includegraphics{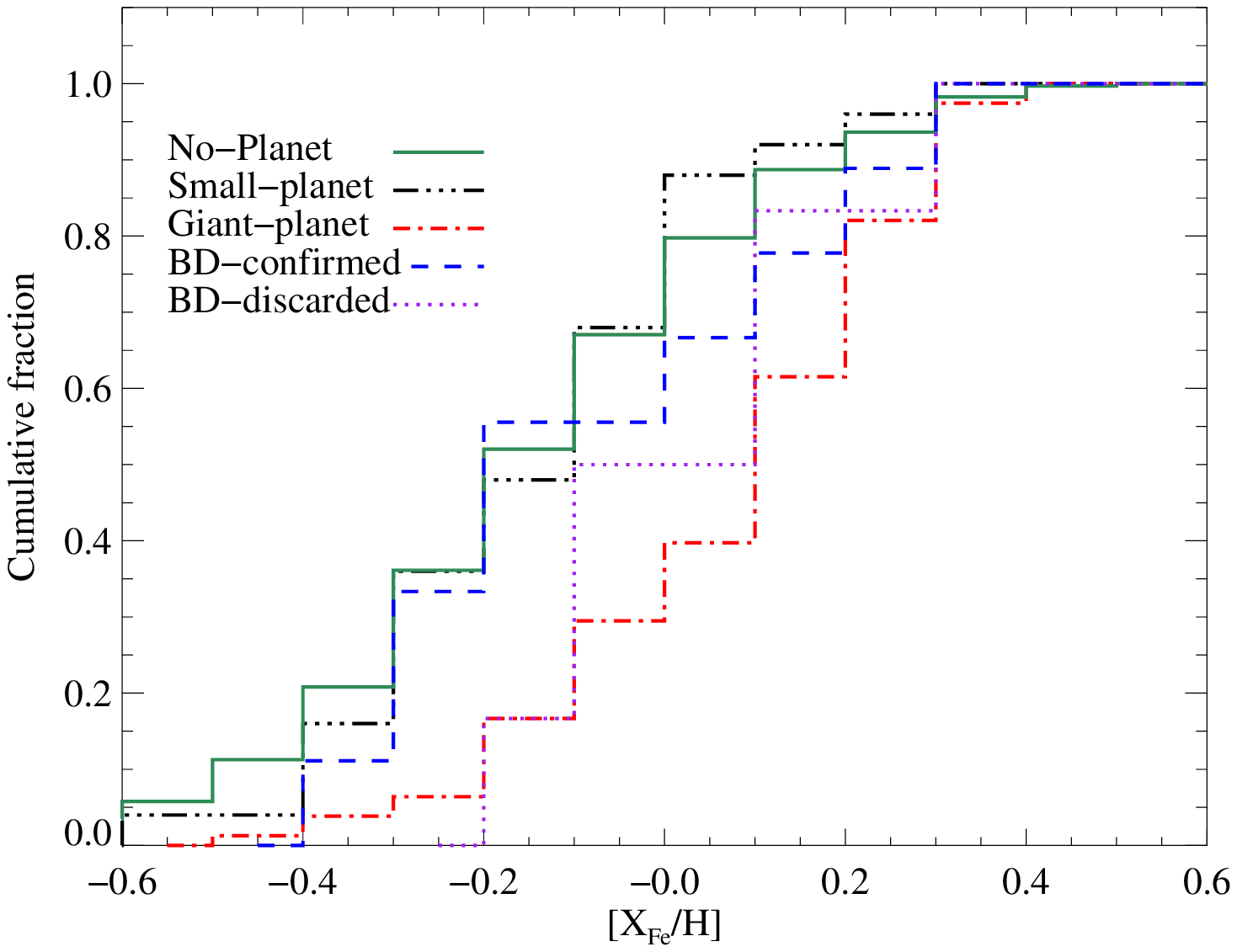}}
\caption{Same as Fig.~\ref{fhalpha} but for Fe-peak element
abundances.} 
\label{fhfe}
\end{figure}

In Figs.~\ref{fhalpha} and~\ref{fhfe} we display the distribution and 
cumulative histograms of the $\alpha$-element abundances (including the
species \ion{Mg}{i}, \ion{Si}{i}, \ion{Ca}{i}, \ion{Ti}{i}), 
$[{\rm X_\alpha/H}]$, and the Fe-peak element abundances (including the
species \ion{Cr}{i}, \ion{Mn}{i}, \ion{Co}{i}, \ion{Ni}{i}), 
$[{\rm X_{\rm Fe}/H}]$. 
Here it appears more clear the fact that confirmed BDs seems to 
behave differently from giant planets. 
Although the sample is small, it may tentatively 
point to a bimodal distribution with two peaks, one at the position of the 
small planet distribution and one at the position of the giant planets. 
However, the mean values of the $[{\rm X_\alpha/H}]$ 
and $[{\rm X_{\rm Fe}/H}]$ abundances are roughly solar.
The $[{\rm X_\alpha/H}]$ and $[{\rm X_{\rm Fe}/H}]$ cumulative histograms 
support the previous statement. Stars with small planets and 
without detected planets go together, whereas the stars with BDs 
exhibit a slightly different behaviour with a later growth of the 
cumulative histogram which resembles that of stars with small planets 
only at [X/H]~$>0.1$~dex. 
The cumulative histogram of stars with giant planets clearly manifest 
a later increase towards high metallicities reaching the saturation 
at [X/H]~$\sim 0.3$~dex. 
We perform a K-S test to statistically evaluate the 
significance of this apparerent different behaviour 
(see Table~\ref{tks}).
This test provides a clear difference between the BD-host sample 
and the GP sample but the SP and NP sample seems to be 
statistically very similar to the BD sample. 
The number of stars with confirmed BDs must be increased in
order to be able to distinguish these populations. 
On the other hand, in Table~\ref{tks} we also show the 
same K-S test for the stars with discarded BDs which are in fact
binaries hosting low-mass M dwarfs. Although again there are only six
stars in this sample, it appears to be statiscally different from all
the samples of GP, SP and NP, especially for the Fe-peak element
abundances. However, the significance is lower than the comparison
with confirmed BD-host stars.

\begin{table}
\caption{Significance of the K-S test of the BD sample}
\label{tks}
\centering
\begin{tabular}{lrrr}
\hline
\hline
BD$^\star$ & SP & GP & NP \\ 
\hline 
 & \multicolumn{3}{c}{$\alpha$-element abundances} \\
\hline 
Discarded & 0.33 & 0.42 &   0.31 \\
Confirmed & 0.87 & 0.08 &   0.83 \\
\hline 
 & \multicolumn{3}{c}{Fe-peak element abundances} \\
\hline 
Discarded & 0.15 & 0.53 &   0.19 \\
Confirmed & 0.72 & 0.13 &   0.77 \\
\hline   
\end{tabular}
\tablefoot{\tablefoottext{$\star$}{K-S test to evaluate the significance
of the behaviour of the abundances [X$_\alpha$/H] and [X$_{\rm Fe}$/H] 
of BD hosts and stars with discarded BDs (see Section~\ref{secdisc}), 
in comparison with stars with small planets (SP), stars with giant 
planets (GP) and stars without detected planets (NP). 
The K-S statistics, with values between 0 and 1, give small values 
to the significance if the cumulative distribution of BD sample is
significantly different from SP, GP or NP samples.}
} 
\end{table}

\subsection{Abundance ratios $\rm [X/H]$ against companion mass}

The $m_C \sin i$ values of the BD candidates in our sample are 
higher than $13\, M_J$ (the star with the lightest BD companion 
analyzed in this work is orbiting HD~211847, $m_C \sin i \sim 19.2\, 
M_J$) but two of the most-massive giant planets of HARPS sample 
exceed this value. 
The boundary between brown dwarfs and giant planets has been 
extensively investigated in the literature, and some works agree with 
the definition that brown dwarfs can burn the deuterium that is 
present when they form, and giant 
planets cannot~\citep[e.g.][]{bur97}. In \citet{bod13}, the 
borderline between giant planets and brown dwarfs is found to depend 
only slightly on different parameters, such as core mass, stellar 
mass, formation location, solid surface density in the protoplanetary 
disc, disc viscosity, and dust opacity. 
More than 50~\% of the initial deuterium is burned for masses 
above $11.6 - 13.6$~$M_J$, in agreement with previous determinations 
that do not take the formation process into account. 
Thus, we keep the mass $\sim 13 M_J$ to distinguish
among giant planets and brown dwarfs.

The metal-content of stars at birth surely affects the formation of 
the planetary-mass companions, but is this also true for stars with BD 
companions? 
To answer this question we depicted in Fig~\ref{fam} the chemical 
element abundances, [X/H], as a function of the minimum mass of the 
most-massive substellar companion, that could be a small planet, a 
giant planet or a brown dwarf according to their $m_C \sin i$ values. 
Qualitatively, these element abundances seem to progressively 
increase with the companion mass from small planets until reaching a 
maximum at about 1~$M_J$ and then slightly decrease when entering in 
the BD regime. 
The scatter in [X/H] may be due to the different intrinsic 
metallicities of the stars at every bin in $m_C \sin i$. 

In Fig.~\ref{fama} we display the mean values of these element 
abundances, [X/H], in bins of $m_C \sin i$. 
The standard deviations from the mean values are different for 
every element, but they stay around $0.15-0.2$~dex.  
These mean element abundances keep roughly constant for small planets 
with masses lower than $0.04\, M_J$.
From this point on, the abundances grow with the companion mass even 
in the giant-planet companion range, from low-mass up to Jupiter-mass 
planets, reaching a maximum in $ \sim 0.8\, M_J$ (see e.g. 
\ion{Si}{i}, \ion{Ti}{i}, \ion{Cr}{i}, and \ion{Ni}{i}). 
For more massive giant-planet companions, these abundances start to 
decrease slowly with the companion mass towards high-mass BD 
companions.
The stars with BD-companion just follow the decreasing trend of the 
stars with giant planets. 

\begin{figure*}
\includegraphics[width=17.5cm]{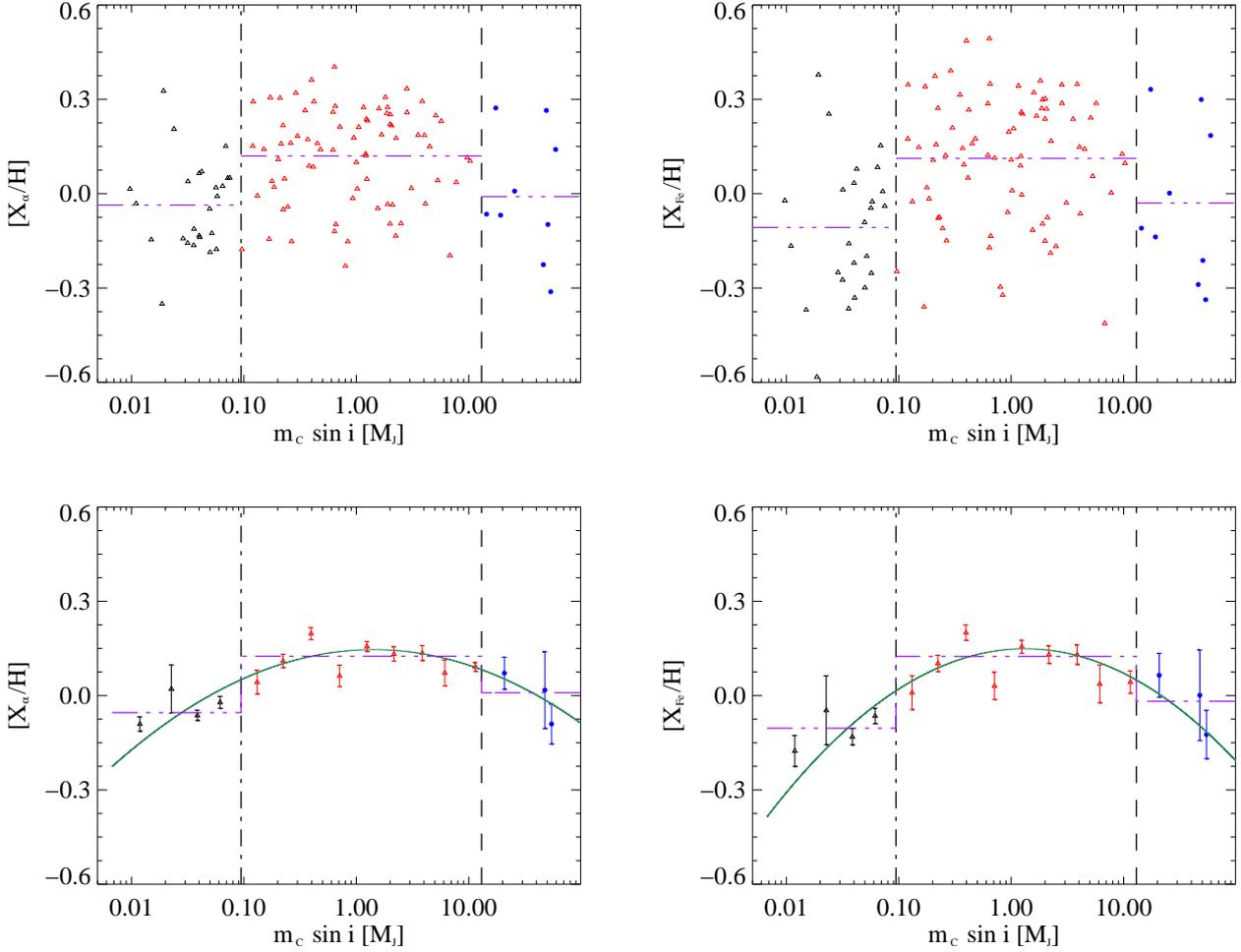}
\caption{Abundance ratios $[{\rm X_\alpha/H}]$ and $[{\rm X_{Fe}/H}]$ 
against minimum mass of the most-massive companion $ m_C \sin i$ 
of $\alpha$-elements (left-top panel) and iron-peak elements 
(right-top panel), for stars with small planets (black empty 
triangles), with giant planets (red empty triangles), with 
confirmed BD companions (blue filled circles). The average values are
shown as violet three-dotted-dashed lines.
Lower panels show the mean abundances of stars with small and giant
planets, and confirmed BD companions, in equal-sized bins 
appropriated for the logarithmic scale of companion masses as in 
Fig.~\ref{fama}. Error bars represents the standard deviation of the mean
divided by the square root of the number of stars in each bin.
The green solid line depicts the parabolic fit of the data, whereas
the violet three-dotted-dashed lines show the zero-order fits.} 
\label{fmamf}
\end{figure*} 

We decided to also use the mean values of the $\alpha$-elements, 
$[{\rm X_\alpha/H}]$ and Fe-peak elements, $[{\rm X_{\rm Fe}/H}]$, of each 
star in these samples. Thus, in Fig.~\ref{fmamf} we display these 
abundances as a function of the minimum mass of the most-massive 
companion, $m_C \sin i$. 
The scatter of these abundances is high due to the 
different global metallicity of different stars.   
The average values of these abundances 
are shown in Table~\ref{tmamf} for the SP, GP and BD samples. One can
see in Fig.~\ref{fmamf} the different levels of these three samples,
although the BD sample shows an average value consistent with the SP
sample within the error bars (see Table~\ref{tmamf}).

\begin{table}
\caption{Mean abundance of SP, GP and BD samples}
\label{tmamf}
\centering
\begin{tabular}{lrrr}
\hline
\hline
Method$^\star$ & SP & GP & BD \\ 
\hline 
 & \multicolumn{3}{c}{$\alpha$-element abundances} \\
\hline 
Average & $-0.04\pm0.03$ & $0.12\pm0.02$ & $-0.01\pm0.07$ \\
Fit     & $-0.05\pm0.01$ & $0.12\pm0.01$ & $0.01\pm0.04$ \\
\hline 
 & \multicolumn{3}{c}{Fe-peak element abundances} \\
\hline 
Average & $-0.11\pm0.04$ & $0.11\pm0.02$ & $-0.03\pm0.08$ \\
Fit     & $-0.10\pm0.02$ & $0.12\pm0.01$ & $-0.02\pm0.05$ \\
\hline   
\end{tabular}
\tablefoot{\tablefoottext{$\star$}{Average values of the  
abundances [X$_\alpha$/H] and [X$_{\rm Fe}$/H] 
of the samples of SP, GP and BDs, depicted as three-dotted-dashed 
lines in top panels of Fig.~\ref{fmamf}, together with the values 
provided by zero-order fits to the weighted average of these element 
abundances, displayed as three-dotted-dashed lines in bottom panels 
of Fig.~\ref{fmamf}. The error bars of the average values show
$\Delta_\sigma=\sigma/\sqrt(N)$, with $\sigma$ equal to the standard 
deviation, and $N$, the number of stars in each sample. 
The fit values have the errors of the coefficients of the zero-order 
functions.}
} 
\end{table}

In the lower panels of Fig.~\ref{fmamf}, we depict the weighted 
average of these abundances at each mass bin and there the trend 
appears more clear. 
We fit a parabolic function, using the IDL routine \textsc{curvefit}, to 
these mean values of the $\alpha$-element 
and Fe-peak element abundances. The peaks of these trends have a 
maximum of abundance at $\sim 0.15 \pm 0.01$~dex and companion mass of 
$m_C \sin i \sim 1.42\pm 0.17$ and $1.32\pm 0.18$~$M_J$, respectively. 
The parabolic fit, with $\chi_\nu^2$ values of 3.2 and 2.3 
respectively, provides a better representation of the average  
abundances than the linear fit (with $\chi_\nu^2\sim 10.5$ and 8.6). 
We perform an F-test to confirm that the parabolic curve fits 
these data set better than the linear fit. 
We use the IDL routine \textsc{mpftest} from Markwardt 
library\footnote{The Markwardt IDL library can be downloaded at:\\
\url{http://cow.physics.wisc.edu/~craigm/idl/idl.html}}. 
The result reveals
a significance level of $1\times10^{-5}$ for $\alpha$-elements 
($\rm F = 45.8$) and $3\times10^{-6}$ for iron-peak elements 
($\rm F = 37.9$), implying that the parabolic fit is significantly 
better than the linear.

We also perform a fit of a three zero-order function of three 
levels describing the SP, GP and BD samples and the values are 
given in Table~\ref{tmamf}, with $\chi_\nu^2$ values of 3.0 and 2.6
for $\alpha$-element and Fe-peak element abundances, respectively.
We compare this 3-step model with the parabolic fit using an F-test,
resulting in $F$ values of -0.5 and 0.6 which gives a significance
level of 0 and $0.5$ for $\alpha$-elements and 
Fe-peak elements, respectively. Therefore, the 3-step model  
provides a similar description of the data than the parabolic fit.
We also check that a 2-step model provides a worse fit than the 3-step
model (with $\chi_\nu^2\sim 3.6$ and 3.1 for $\alpha$-elements and 
Fe-peak elements, respectively).

Finally, \citet{sah11} noticed that there is a lack of BD companions 
with masses in the range $m_C \sin i \sim $~35--55~$M_J$. 
More recently, \citet{mag14} have collected the known BD companions 
from different studies and confirm this gap for stars with for 
periods shorter than 100 days. 
Although the statistics may be still poor these authors
suggest that BD companions below this gap, i.e. with 
$m_C \sin i < 42$~$M_J$, may have formed in protoplanetary disks
as giant planets, probably through the disk instability-fragmentation
mechanism~\citep{bos97,sta09}, 
whereas BD companions with $m_C \sin i > 42$~$M_J$ may have 
formed by molecular cloud fragmentation as 
stars~\citep{pad04,hen08}. 

In bottom panels of Fig.~\ref{fmamf} we may see that on average 
stars with BDs at masses $m_C \sin i$ below 42~$M_J$ have 
higher $\alpha$-element and Fe-peak element abundances than stars 
above this mass. 
In fact, stars with massive BD companions have more similar 
abundances to those of stars without planets. 
This might tentatively support the above statement with the low-mass 
BD companions being formed in protoplanetary disks as giant planets, 
and the high-mass BDs being formed by cloud fragmentation as stars.

%__________________________________________________________________

\section{Conclusions\label{sec7}}

We have analyzed a subsample of stars with candidate BD companions 
from the CORALIE radial velocity survey. We derive chemical abundances 
of several elements including $\alpha$-elements and Fe-peak elements. 
A comparison with the chemical abundances of stars with giant planets 
shows that BD-host stars seem to behave differently.
In particular, we compute the abundance histograms [X$_\alpha$/H] and 
[X$_{\rm Fe}$/H], revealing a mean abundance at about solar for the 
BD-host sample whereas for stars without planets (NP) and with small 
planets (SP) remains at $-0.1$~dex, and for stars with giant planets 
(GP) at roughly$+0.10$~dex.
The cumulative histograms of [X$_\alpha$/H] and
[X$_{\rm Fe}$/H] abundances exhibit the same situation, with the 
stars without planets and with small planets going together, 
similarly to the stars with BDs. 
However, the stars with giant planets reach a later 
saturation at [X/H]~$\sim 0.3$~dex. 
A Kosmogorov-Smirnov (K-S) test does not show a statisticallly 
significant difference between the cumulative distribution of 
SP, NP and BD samples, but clearly separates the GP and BD samples.

Finally, we depict the [X$_\alpha$/H] and [X$_{\rm Fe}$/H] abundances 
versus the minimum mass of the most-massive substellar companion, 
$m_C \sin i$, and we find a peak of these element abundances for a 
companion mass $m_C \sin i \sim 1.3-1.4\, M_J$, with the abundances 
growing with the companion mass from small planets to Jupiter-like 
planets and after decresing towards massive BD companions. 
A 3-step model also provides a similar description of the data
with no statistically significant difference with the parabolic
model.
Recently, \citet{sah11} and \citet{mag14} have suggested that the 
formation mechanism may be different for BD companion below and 
above 42~$M_J$. We find that BDs below this mass tend to have higher 
abundances than those above this mass, which may support this 
conclusion and BDs with $m_C \sin i < 42$~$M_J$ may form by disk 
instability-fragmentation whereas high-mass BD may form as stars 
by cloud fragmentation.

\section{Acknowledgments}
D.M.S. is grateful to the Spanish Ministry of Education, Culture and
Sport for the financial support from a collaboration grant, and to 
the PhD contract funded by Fundaci\'on La Caixa.  
J.I.G.H. and G.I. acknowledge financial support from the 
Spanish Ministry project MINECO AYA2011-29060, and J.I.G.H.
also from the Spanish Ministry of Economy and Competitiveness
(MINECO) under the 2011 Severo Ochoa Program MINECO SEV-2011-0187. 
N.C.S. acknowledges the support by the European Research 
Council/European Community under the FP7 through Starting Grant
agreement number 239953, and the support in the form of a Investigador 
FCT contract funded by Funda\c{c}\~ao para a Ci\^encia e a Tecnologia 
(FCT) /MCTES (Portugal) and POPH/FSE (EC).
This research has made use of the IRAF facilities, and the SIMBAD 
database, operated at the CDS, Strasbourg, France.

\bibliographystyle{aa}
\bibliography{planetBDaa}

\clearpage

\appendix
\section{Tables of chemical abundaces}
In this appendix we provide all the tables containing the element
abundances of the 17 stars within the CORALIE sample, the first 
13 stars with BD-companion candidates (for more details see the text
of Section~\ref{sec3}, 7 confirmed BDs and 6 discarded BDs, and 
four stars as comparison sample.

\begin{table*}
\caption{Abundances of the CORALIE stellar sample. Error
bars show the standard deviation from individual measurements from
different lines.   
For elements with a single line, we adopt an error bar of $0.10$.}
\label{tabu1}
\centering
\begin{tabular}{lrrrrr}
\hline
\hline 
Star  & $\rm [NaI/H]$ & $\rm [MgI/H]$ & $\rm [AlI/H]$ & $\rm [SiI/H]$ & 
$\rm [CaI/H]$ \\
\hline
\hline
\\
  HD3277  & $ 0.02  \pm 0.08  $ & $-0.07  \pm 0.05  $ & $-0.01  \pm 0.10  $ & $-0.09  \pm 0.04  $ & $-0.04  \pm 0.06   $ \\
  HD4747  & $-0.24  \pm 0.01  $ & $-0.23  \pm 0.04  $ & $-0.27  \pm 0.04  $ &  $-0.24  \pm 0.03  $ & $-0.23  \pm 0.06	$ \\
 HD43848  & $ 0.48  \pm 0.08  $ & $ 0.44  \pm 0.19  $ & $ 0.43  \pm 0.06  $ &	 $ 0.21  \pm 0.16  $ & $ 0.28  \pm 0.17   $  \\
 HD52756  & $ 0.31  \pm 0.06  $ & $ 0.13  \pm 0.06  $ & $ 0.23  \pm 0.04  $ &  $ 0.11  \pm 0.09  $ & $ 0.08  \pm 0.12	$  \\
 HD74014  & $ 0.37  \pm 0.01  $ & $ 0.30  \pm 0.04  $ & $ 0.31  \pm 0.04  $ &	 $ 0.23  \pm 0.08  $ & $ 0.23  \pm 0.07   $  \\
 HD89707  & $-0.36  \pm 0.18  $ & $-0.40  \pm 0.23  $ & $-0.48  \pm 0.14  $ &	 $-0.34  \pm 0.11  $ & $-0.28  \pm 0.08   $  \\
HD154697  & $ 0.09  \pm 0.04  $ & $ 0.14  \pm 0.03  $ & $ 0.16  \pm 0.03  $ &	 $ 0.08  \pm 0.06  $ & $ 0.11  \pm 0.05   $  \\
HD164427A & $ 0.22  \pm 0.04  $ & $ 0.16  \pm 0.05  $ & $ 0.12  \pm 0.10  $ &	 $ 0.15  \pm 0.04  $ & $ 0.13  \pm 0.09   $  \\
HD167665  & $-0.12  \pm 0.08  $ & $-0.13  \pm 0.08  $ & $-0.37  \pm 0.01  $ &	 $-0.09  \pm 0.13  $ & $-0.10  \pm 0.22   $  \\
 HD17289  & $-0.03  \pm 0.08  $ & $-0.12  \pm 0.03  $ & $-0.16  \pm 0.24  $ &	 $-0.23  \pm 0.24  $ & $-0.09  \pm 0.08   $  \\
HD189310  & $ 0.03  \pm 0.05  $ & $-0.01  \pm 0.09  $ & $ 0.07  \pm 0.03  $ &	 $-0.03  \pm 0.06  $ & $-0.02  \pm 0.17   $  \\
HD211847  & $-0.18  \pm 0.06  $ & $-0.03  \pm 0.09  $ & $-0.05  \pm 0.06  $ &	 $-0.12  \pm 0.06  $ & $-0.06  \pm 0.10   $  \\
 HD30501  & $ 0.00  \pm 0.06  $ & $-0.02  \pm 0.02  $ & $ 0.03  \pm 0.03  $ &	 $-0.07  \pm 0.06  $ & $-0.02  \pm 0.14   $  \\
 HD74842  & $ 0.03  \pm 0.11  $ & $-0.18  \pm 0.10  $ & $-0.01  \pm 0.11  $ &	 $-0.10  \pm 0.06  $ & $ 0.01  \pm 0.07   $  \\
 HD94340  & $ 0.20  \pm 0.05  $ & $ 0.10  \pm 0.06  $ & $ 0.18  \pm 0.10  $ &	 $ 0.14  \pm 0.05  $ & $ 0.13  \pm 0.07   $  \\
HD112863  & $-0.13  \pm 0.04  $ & $-0.23  \pm 0.07  $ & $-0.16  \pm 0.04  $ &	 $-0.12  \pm 0.08  $ & $-0.09  \pm 0.06   $  \\
HD206505  & $ 0.26  \pm 0.10  $ & $ 0.17  \pm 0.09  $ & $ 0.24  \pm 0.10  $ &	 $ 0.11  \pm 0.06  $ & $ 0.05  \pm 0.09   $  \\
\hline   
\end{tabular}
\end{table*}

\begin{table*}
\caption{Abundances of the CORALIE stellar sample. Error
bars show the standard deviation from individual measurements from
different lines.   
For elements with a single line, we adopt an error bar of $0.10$.}
\label{tabu2}
\centering
\begin{tabular}{lrrrrr}
\hline
\hline    
Star &    $\rm [ScI/H]$ &    $\rm [ScII/H]$  &   $\rm [TiI/H]$  &   $\rm [TiII/H]$   &  $\rm [VI/H]$  
\\
\hline
\hline
\\
  HD3277  & $-0.03  \pm 0.04  $ & $-0.15  \pm 0.07  $ &    $-0.03  \pm 0.06  $ & $-0.13  \pm 0.04  $ &    $ 0.03  \pm 0.06   $  \\
  HD4747  & $-0.27  \pm 0.15  $ & $-0.29  \pm 0.05  $ &    $-0.20  \pm 0.04  $ & $-0.22  \pm 0.04  $ &    $-0.14  \pm 0.07   $  \\
 HD43848  & $ 0.54  \pm 0.19  $ & $ 0.19  \pm 0.15  $ &    $ 0.36  \pm 0.11  $ & $ 0.27  \pm 0.16  $ &    $ 0.68  \pm 0.17   $  \\
 HD52756  & $ 0.30  \pm 0.16  $ & $ 0.07  \pm 0.10  $ &    $ 0.24  \pm 0.13  $ & $ 0.09  \pm 0.12  $ &    $ 0.49  \pm 0.19   $  \\
 HD74014  & $ 0.33  \pm 0.10  $ & $ 0.27  \pm 0.14  $ &    $ 0.30  \pm 0.07  $ & $ 0.21  \pm 0.08  $ &    $ 0.42  \pm 0.08   $  \\
 HD89707  & $-0.07  \pm 0.19  $ & $-0.25  \pm 0.08  $ &    $-0.23  \pm 0.16  $ & $-0.26  \pm 0.05  $ &    $-0.36  \pm 0.21   $  \\
HD154697  & $ 0.16  \pm 0.10  $ & $ 0.12  \pm 0.04  $ &    $ 0.13  \pm 0.05  $ & $ 0.09  \pm 0.09  $ &    $ 0.19  \pm 0.07   $  \\
HD164427A & $ 0.11  \pm 0.12  $ & $ 0.24  \pm 0.03  $ &    $ 0.14  \pm 0.07  $ & $ 0.21  \pm 0.03  $ &    $ 0.16  \pm 0.04   $  \\
HD167665  & $-0.45  \pm 0.10  $ & $ 0.09  \pm 0.30  $ &    $-0.07  \pm 0.07  $ & $-0.04  \pm 0.05  $ &    $-0.11  \pm 0.19   $  \\
 HD17289  & $-0.03  \pm 0.04  $ & $-0.15  \pm 0.14  $ &    $-0.06  \pm 0.16  $ & $-0.12  \pm 0.03  $ &    $ 0.03  \pm 0.08   $  \\
HD189310  & $ 0.10  \pm 0.17  $ & $-0.02  \pm 0.09  $ &    $ 0.09  \pm 0.08  $ & $ 0.02  \pm 0.05  $ &    $ 0.30  \pm 0.14   $  \\
HD211847  & $-0.11  \pm 0.05  $ & $-0.19  \pm 0.05  $ &    $-0.06  \pm 0.05  $ & $-0.15  \pm 0.10  $ &    $-0.08  \pm 0.04   $  \\
 HD30501  & $ 0.03  \pm 0.12  $ & $-0.15  \pm 0.07  $ &    $ 0.04  \pm 0.07  $ & $-0.11  \pm 0.06  $ &    $ 0.24  \pm 0.14   $  \\
 HD74842  & $-0.20  \pm 0.31  $ & $-0.16  \pm 0.05  $ &    $-0.01  \pm 0.14  $ & $-0.08  \pm 0.02  $ &    $ 0.03  \pm 0.18   $  \\
 HD94340  & $ 0.21  \pm 0.09  $ & $ 0.17  \pm 0.04  $ &    $ 0.11  \pm 0.07  $ & $ 0.09  \pm 0.02  $ &    $ 0.18  \pm 0.07   $  \\
HD112863  & $-0.13  \pm 0.16  $ & $-0.19  \pm 0.08  $ &    $-0.04  \pm 0.05  $ & $-0.13  \pm 0.11  $ &    $ 0.05  \pm 0.07   $  \\
HD206505  & $ 0.20  \pm 0.12  $ & $ 0.11  \pm 0.05  $ &    $ 0.18  \pm 0.06  $ & $ 0.11  \pm 0.09  $ &    $ 0.39  \pm 0.12   $  \\
\hline  
\end{tabular}
\end{table*}

\begin{table*}
\caption{Abundances of the CORALIE stellar sample. Error
bars show the standard deviation from individual measurements from
different lines.   
For elements with a single line, we adopt an error bar of $0.10$.}
\label{tabu3}
\centering
\begin{tabular}{lrrrrr}
\hline
\hline  
Star  &    $\rm [CrI/H]$  &  $\rm [CrII/H]$  &   $\rm [MnI/H]$  &   $\rm [CoI/H]$   &  $\rm [NiI/H]$
\\
\hline
\hline
\\
  HD3277   & $-0.07  \pm 0.06  $ & $-0.16  \pm 0.01  $ & $-0.06  \pm 0.04  $ &    $-0.11  \pm 0.05  $ & $-0.09  \pm 0.04   $  \\
  HD4747   & $-0.26  \pm 0.06  $ & $-0.24  \pm 0.04  $ & $-0.29  \pm 0.08  $ &    $-0.29  \pm 0.04  $ & $-0.32  \pm 0.03   $  \\
 HD43848   & $ 0.27  \pm 0.11  $ & $ 0.13  \pm 0.20  $ & $ 0.30  \pm 0.32  $ &    $ 0.50  \pm 0.20  $ & $ 0.25  \pm 0.10   $  \\
 HD52756   & $ 0.14  \pm 0.09  $ & $ 0.02  \pm 0.12  $ & $ 0.14  \pm 0.06  $ &    $ 0.31  \pm 0.10  $ & $ 0.14  \pm 0.07   $  \\
 HD74014   & $ 0.23  \pm 0.03  $ & $ 0.16  \pm 0.07  $ & $ 0.35  \pm 0.03  $ &    $ 0.34  \pm 0.05  $ & $ 0.27  \pm 0.06   $  \\
 HD89707   & $-0.32  \pm 0.19  $ & $-0.34  \pm 0.07  $ & $-0.31  \pm 0.61  $ &    $-0.29  \pm 0.11  $ & $-0.43  \pm 0.18   $  \\
HD154697   & $ 0.09  \pm 0.06  $ & $ 0.04  \pm 0.03  $ & $ 0.10  \pm 0.12  $ &    $ 0.13  \pm 0.05  $ & $ 0.11  \pm 0.05   $  \\
HD164427A  & $ 0.09  \pm 0.05  $ & $ 0.16  \pm 0.06  $ & $ 0.17  \pm 0.02  $ &    $ 0.14  \pm 0.03  $ & $ 0.19  \pm 0.04   $  \\
HD167665   & $-0.17  \pm 0.10  $ & $-0.07  \pm 0.08  $ & $-0.21  \pm 0.11  $ &    $-0.27  \pm 0.08  $ & $-0.20  \pm 0.13   $  \\
 HD17289   & $-0.11  \pm 0.10  $ & $-0.05  \pm 0.08  $ & $-0.16  \pm 0.09  $ &    $-0.17  \pm 0.08  $ & $-0.19  \pm 0.07   $  \\
HD189310   & $-0.05  \pm 0.09  $ & $ 0.00  \pm 0.08  $ & $ 0.01  \pm 0.06  $ &    $ 0.04  \pm 0.02  $ & $-0.00  \pm 0.06   $  \\
HD211847   & $-0.09  \pm 0.05  $ & $-0.14  \pm 0.06  $ & $-0.11  \pm 0.05  $ &    $-0.19  \pm 0.05  $ & $-0.15  \pm 0.04   $  \\
 HD30501   & $-0.03  \pm 0.09  $ & $-0.14  \pm 0.08  $ & $-0.08  \pm 0.16  $ &    $-0.04  \pm 0.05  $ & $-0.09  \pm 0.07   $  \\
 HD74842   & $-0.06  \pm 0.08  $ & $-0.14  \pm 0.06  $ & $-0.09  \pm 0.06  $ &    $-0.18  \pm 0.06  $ & $-0.13  \pm 0.08   $  \\
 HD94340   & $ 0.09  \pm 0.07  $ & $ 0.10  \pm 0.09  $ & $ 0.13  \pm 0.05  $ &    $ 0.14  \pm 0.06  $ & $ 0.12  \pm 0.05   $  \\
HD112863   & $-0.09  \pm 0.07  $ & $-0.13  \pm 0.16  $ & $-0.15  \pm 0.11  $ &    $-0.18  \pm 0.08  $ & $-0.18  \pm 0.08   $  \\
HD206505   & $ 0.12  \pm 0.08  $ & $ 0.06  \pm 0.07  $ & $ 0.18  \pm 0.08  $ &    $ 0.21  \pm 0.08  $ & $ 0.15  \pm 0.07   $  \\
\hline  
\end{tabular}
\end{table*}  

\clearpage %Para forzar a limpiar y poner todo las demas imagenes detras

\begin{table*}

\caption[]{Minimum mass of the most-massive substellar companion of stars 
in the HARPS sample.}
\centering
  \begin{tabular}{lc}
  \hline\hline
   Star name & Companion Mass ($M_J$)\\
    \hline\hline
215152 &  0.010 \\    \hline
85512 &  0.011 \\    \hline
20794 &  0.015 \\    \hline
39194 &  0.019 \\    \hline
154088 &  0.019 \\    \hline
1461 &  0.024 \\    \hline
40307 &  0.029 \\    \hline
189567 &  0.032 \\    \hline
93385 &  0.032 \\    \hline
136352 &  0.036 \\    \hline
13808 &  0.036 \\    \hline
45184 &  0.040 \\    \hline
96700 &  0.040 \\    \hline
4308 &  0.041 \\    \hline
20003 &  0.042 \\    \hline
20781 &  0.050 \\    \hline
102365 &  0.050 \\    \hline
31527 &  0.052 \\    \hline
51608 &  0.056 \\    \hline
90156 &  0.057 \\    \hline
69830 &  0.058 \\    \hline
21693 &  0.065 \\    \hline
16417 &  0.069 \\    \hline
115617 &  0.072 \\    \hline
192310 &  0.075 \\    \hline
38858 &  0.096 \\    \hline
157172 &  0.120 \\    \hline
134606 &  0.121 \\    \hline
85390 &  0.132 \\    \hline
134060 &  0.151 \\    \hline
150433 &  0.168 \\    \hline
102117 &  0.172 \\    \hline
117618 &  0.178 \\    \hline
104067 &  0.186 \\    \hline
10180 &  0.203 \\    \hline
107148 &  0.210 \\    \hline
16141 &  0.215 \\    \hline
137388 &  0.223 \\    \hline
126525 &  0.224 \\    \hline
168746 &  0.230 \\    \hline
215456 &  0.246 \\    \hline
108147 &  0.261 \\    \hline
204941 &  0.266 \\    \hline
7199 &  0.290 \\    \hline
101930 &  0.300 \\    \hline
47186 &  0.351 \\    \hline
93083 &  0.370 \\    \hline
63454 &  0.380 \\    \hline
83443 &  0.400 \\    \hline
208487 &  0.413 \\    \hline
75289 &  0.420 \\    \hline
212301 &  0.450 \\    \hline
2638 &  0.480 \\    \hline
27894 &  0.620 \\    \hline
330075 &  0.620 \\    \hline
181433 &  0.640 \\    \hline
\end{tabular}
\end{table*}

\begin{table*}
\caption[]{Minimum mass of the most-massive substellar companion of stars 
in the HARPS sample.}
\label{tplm}
\centering
\begin{tabular}{lc}
\hline\hline
Star HD number & Companion Mass ($M_J$)\\
\hline\hline
63765 &  0.640 \\    \hline
216770 &  0.650 \\    \hline
45364 &  0.658 \\    \hline
209458 &  0.714 \\    \hline
4208 &  0.800 \\    \hline
114729 &  0.840 \\    \hline
10647 &  0.930 \\    \hline
179949 &  0.950 \\    \hline
114783 &  1.000 \\    \hline
130322 &  1.020 \\    \hline
52265 &  1.050 \\    \hline
100777 &  1.160 \\    \hline
147513 &  1.210 \\    \hline
121504 &  1.220 \\    \hline
210277 &  1.230 \\    \hline
114386 &  1.240 \\    \hline
216435 &  1.260 \\    \hline
22049 &  1.550 \\    \hline
134987 &  1.590 \\    \hline
19994 &  1.680 \\    \hline
160691 &  1.814 \\    \hline
73256 &  1.870 \\    \hline
20782 &  1.900 \\    \hline
190647 &  1.900 \\    \hline
7449 &  2.000 \\    \hline
70642 &  2.000 \\    \hline
82943 &  2.010 \\    \hline
117207 &  2.060 \\    \hline
159868 &  2.100 \\    \hline
65216 &  2.240 \\    \hline
17051 &  2.260 \\    \hline
23079 &  2.500 \\    \hline
66428 &  2.820 \\    \hline
196050 &  2.830 \\    \hline
221287 &  3.090 \\    \hline
204313 &  3.550 \\    \hline
92788 &  3.860 \\    \hline
169830 &  4.040 \\    \hline
166724 &  4.120 \\    \hline
213240 &  4.500 \\    \hline
142022A &  5.100 \\    \hline
142 &  5.300 \\    \hline
28185 &  5.700 \\    \hline
111232 &  6.800 \\    \hline
222582 &  7.750 \\    \hline
141937 &  9.700 \\    \hline
39091 & 10.300 \\    \hline
162020 & 14.400 \\    \hline
202206 & 17.400 \\    \hline
\end{tabular}
\end{table*}

\clearpage %Para forzar a limpiar y poner todo las demas imagenes detras

\section{Individual element abundance figures and histograms}

\begin{figure*}
\centering
\includegraphics[width=17.5cm]{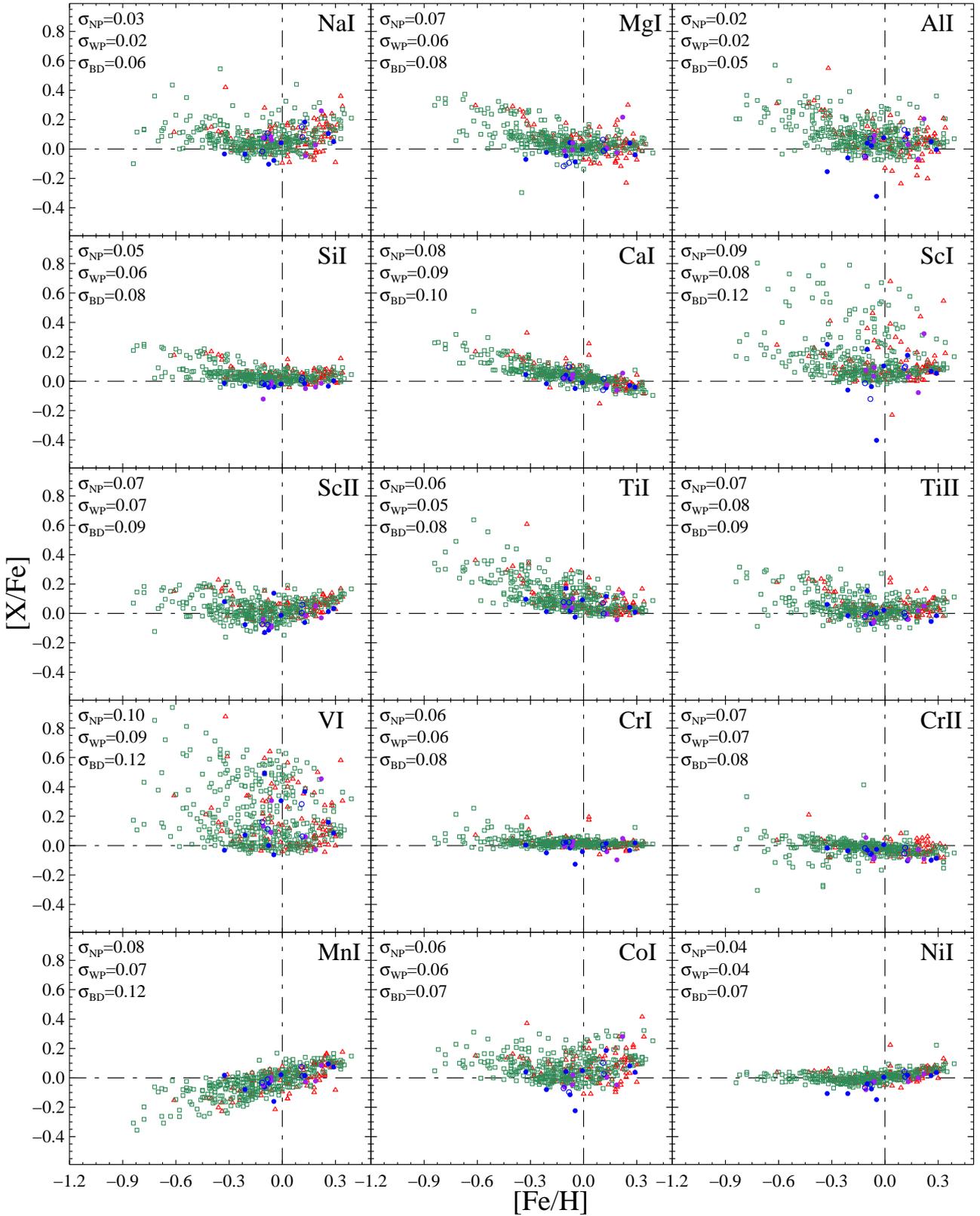}
\caption{Element abundance ratios $[{\rm X/Fe}]$ 
against [Fe/H] for the samples of stars without planets (green empty squares),
stars with planets (red empty triangles), BD-host stars (blue filled circles),
stars without known BD companions from the CORALIE sample (blue empty circles), 
and stars with discarded BD candidates (violet filled circles). 
At the top-left corner of each panel, the mean standard deviation of the 
abundance measurements are shown for the stars without planetary companion 
($\sigma_{\rm NP}$), stars with planets ($\sigma_{\rm WP}$) and stars with
confirmed BD companions ($\sigma_{\rm BD}$)
respectively.} 
\label{ftrend}
\end{figure*} 

\begin{figure*}
\includegraphics[width=17.5cm]{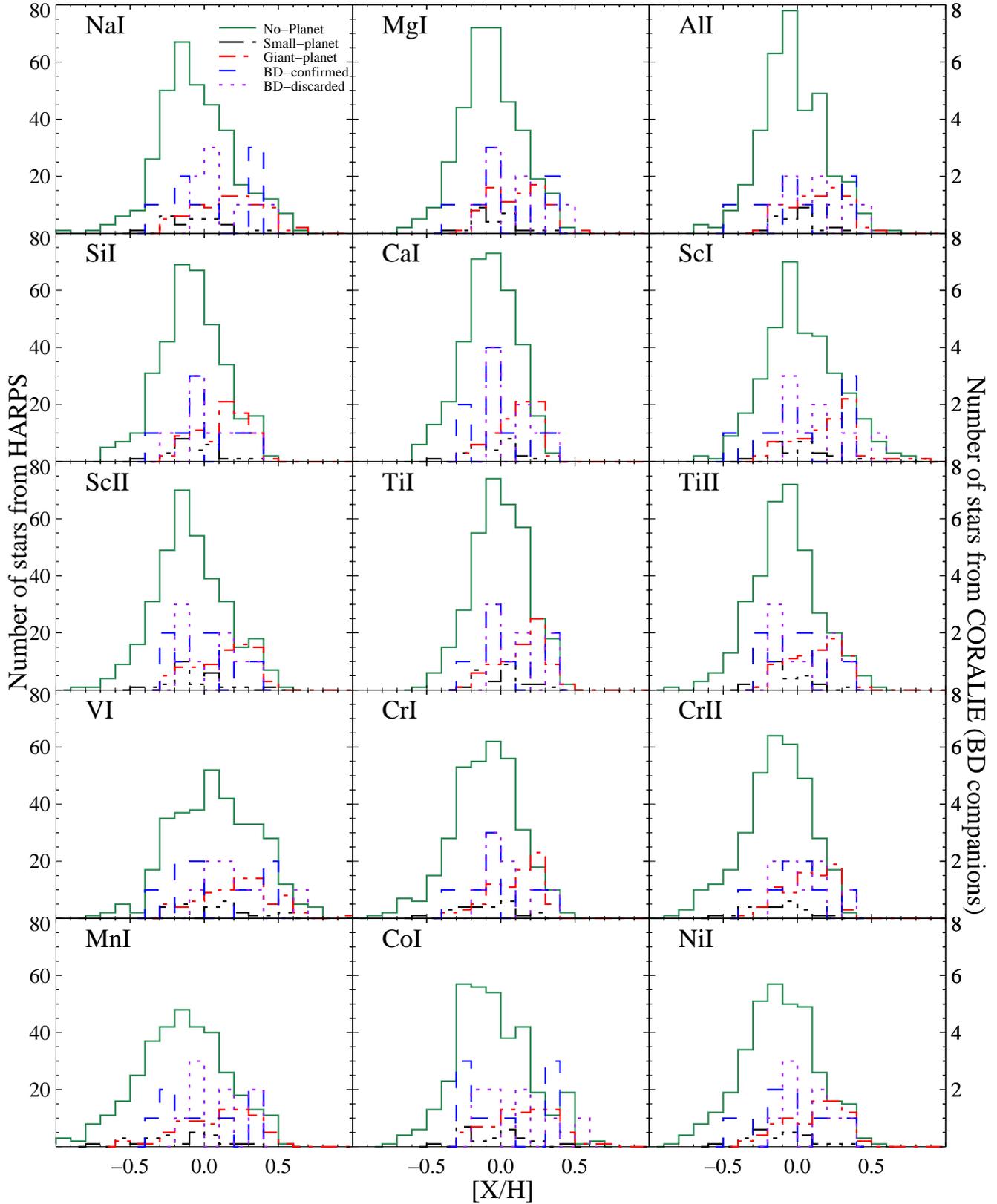}
\caption{Histogram of element abundances, $[{\rm X/H}]$, 
of the samples of stars without planets (green continuum line), 
stars with low-mass planets (black dashed-double-dotted line), stars with 
giant planets (red dashed-dotted line), stars with confirmed BD 
companions (blue dashed line) and stars with discarded BD candidates 
(violet dotted line). 
The left y-axis of the top panel is labeled with the number of stars 
with and without small and giant planets whereas the right y-axis 
shows the number of stars with BD-companion candidates.} 
\label{fha}
\end{figure*} 

\begin{figure*}
\includegraphics[width=17.5cm]{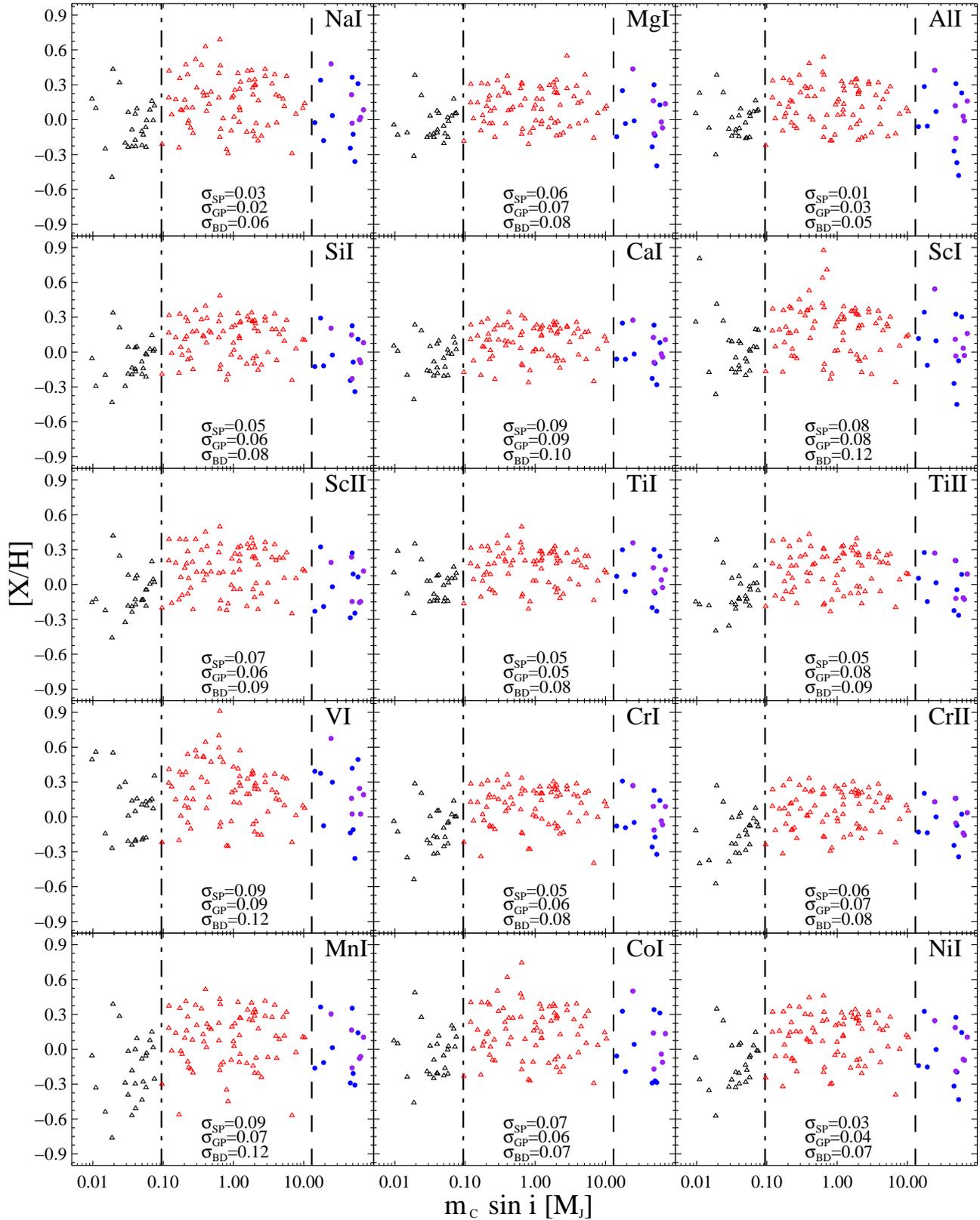}
\caption{Element abundances, $\rm [X/H]$, against the minimum mass of the
most-massive companion, $m_C \sin i$. 
Black empty triangles refer to stars with small planets, red
empty triangles to stars with giant planets, and blue filled circles
to stars with confirmed BDs, and violet filled circles to stars with 
discarded BD candidates. 
At the bottom of each panel, standard deviations of the mean 
element abundances are shown for stars without planets
($\sigma_{\rm NP}$), stars with planets ($ \sigma_{\rm WP}$) and stars with 
confirmed BD companions ($\sigma_{\rm BD}$), respectively.} 
\label{fam}
\end{figure*} 

\begin{figure*}
\includegraphics[width=17.5cm]{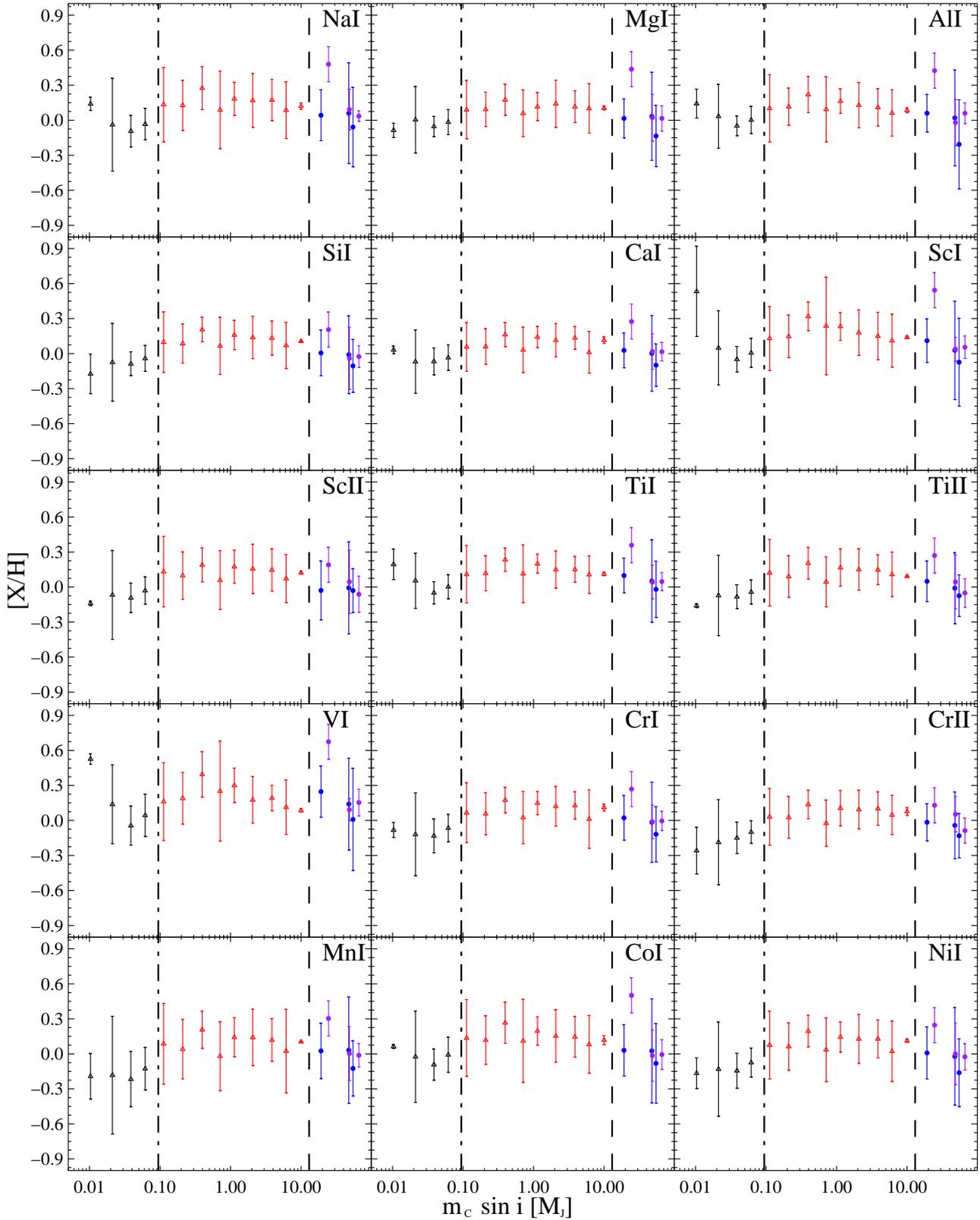}
\caption{Mean element abundances versus the minimum mass of the
most-massive companion, $m_C \sin i$, computed in equal-size bins 
appropriated for the logarithmic scale of companion masses. 
Error bars represent the standard deviation of the mean.} 
\label{fama}
\end{figure*} 

\end{document}